\documentstyle[aps,prd,eqsecnum,preprint,tighten,floats,epsf,rotate]{revtex}

\begin{document}

\draft  

\title{A calculation of the $B_{B}^{}$ parameter \\
       in the static limit }

\author{Joseph Christensen, Terrence Draper and Craig
        McNeile~\footnote{Now at Department of Physics, University of
        Utah, Salt Lake City, UT 84112}}
\address{Department of Physics and Astronomy,
         University of Kentucky, Lexington, KY 40506, USA}

\maketitle

\begin{abstract}

We calculate the $B_{B}^{}$ parameter, relevant for
$\overline{B}^0${\bf --}$B^0$ mixing, from a lattice gauge theory
simulation at $\beta = 6.0$.  The bottom quarks are simulated in the
static theory, the light quarks with Wilson fermions.  Improved
smearing functions produced by a variational technique, {\sc most},
are used to reduce statistical errors and minimize excited-state
contamination of the ground-state signal.  We obtain
$B_B^{}(4.33\,{\rm GeV}) =
0.98^{+4}_{-4}\,{\rm(statistical)}^{+3}_{-18}\,{\rm(systematic)}$
which corresponds to $\widehat{B}_B^{}$ =
$1.40^{+6}_{-6}\,{\rm(statistical)}^{+4}_{-26}\,{\rm(systematic)}$
for the one-loop renormalization-scheme-independent parameter.  The
systematic errors include the uncertainty due to alternative (less
favored) treatments of the perturbatively-calculated mixing
coefficients; this uncertainty is at least as large as residual
differences between Wilson-static and clover-static results.  Our
result agrees with extrapolations of results from relativistic
(Wilson) heavy quark simulations.

\end{abstract}

%
\pacs{12.38.Gc,14.40.Nd,12.39.Hg,12.38.Bx}

\section{Introduction}

The experimental observation of $\overline{B}^0${\bf --}$B^0$ mixing
allows, in principle, the extraction of the $\mid \! V_{td} \! \mid$
CKM matrix element~\cite{Rosner92a,Soni96a}.  The over-determination
of the CKM matrix is a high-precision test of the standard model of
particle physics and is regarded as a potential harbinger of new
physics.  The dominant uncertainty in the extraction of $\mid \!
V_{td}\! \mid$ from experimental measurements is due to theoretical
factors from non-perturbative QCD\@.  The key factor is $B_{B}^{}
f_{B}^{2}$, where $f_{B}^{}$ is the $B$-meson semi-leptonic decay
constant and $B_{B}^{}$ is the ``bag constant'' for the $B$-meson,
defined as the ratio of the matrix element of the operator relevant
for the mixing to its value in the vacuum-saturation approximation
(VSA).

There have been a large number of lattice gauge theory simulations
which have calculated the $f_{B}^{}$ decay constant; however, much
less work has been done on the calculation of the $B_{B}^{}$
parameter.  The earliest result~\cite{Bernard88a} suggested that the
VSA works quite well; this result was unanticipated and is quite
non-trivial, as was reiterated by Soni~\cite{Soni96a}.  Later results
by other groups are surprisingly scattered, with significant
disagreement in some cases~\cite{Flynn97a} and with some results
markedly different than that suggested by VSA\@.  Here we argue that,
in fact, most raw lattice data are consistent with VSA (including
ours which are quite precise due to the use of improved smearing
functions) and that groups differ due to their choices of how to
relate these to the full-theory continuum value.  We argue that
although large systematic uncertainties remain due to unknown
higher-order contributions in the mixing coefficients, it is possible
to formulate the calculation in a way which is stable against changes
in normalization (such as tadpole improvement).  Our result is in
accord with VSA and is also in agreement with the large-mass
extrapolation of calculations~\cite{Soni96a} which use relativistic,
rather than static, heavy quarks.

Some of the first attempts at simulating the static theory calculated
both the decay constant and the $B_{B}^{}$
parameter~\cite{Allton91a,Eichten90a}.  However, the required
perturbative matching coefficients were not known; these have since
been computed by Flynn {\it et al.}~\cite{Flynn91a}.  Their analysis
showed that additional operators, not included in the first
simulations, are required to estimate the $B_{B}^{}$ parameter.

Until recently, the simulation of the static theory was problematic
because of excited-state contamination of the ground-state
signal~\cite{Lepage92a,Bernard94a,Allton96a}.  The development of
variational techniques~\cite{Duncan94c,Draper94a} has finally allowed
a reliable extraction of the decay constant.  In this paper, we use a
modern variational technique~\cite{Draper94a} to obtain accurate
estimates of the lattice matrix elements and combine these with the
mixing coefficients to calculate the static $B_{B}^{}$ parameter.  At
two conferences~\cite{Draper96a,Christensen97a}, we have reported
preliminary results for the value of $B_{B}^{}$ from this simulation.

Sec.~\ref{Sec:NumTec} outlines the method of extracting the relevant
matrix elements from lattice correlation functions;
Sec.~\ref{Sec:NumRes} summarizes our numerical results.
Sec.~\ref{Sec:PerMat} contains a summary of the perturbative-matching
techniques which, rather explicitly, details our preferred way of
organizing the calculation; we argue that our method reduces
systematic errors in the matching coefficients which are then
estimated in Sec.~\ref{Sec:SysErr}.  In Sec.~\ref{Sec:Compare}, a
comparison is made to other groups as an illustration of the
differences in the methods discussed in Sec.~\ref{Sec:SysErr}.  The
conclusion follows as Sec.~\ref{Sec:Con}.

\section{Numerical techniques}
\label{Sec:NumTec}

The static-light $B_{B}^{}$ parameter is obtained from a combination
of two- and three-point hadronic correlation functions.  The required
three-point function is
\begin{equation}
C_{3,X}(t_1, t_2 )
  =   \sum_{\vec{x}_1} \sum_{\vec{x}_2}
      \left< \, 0
      \left| {\cal T}
      \left( \chi (t_1,\vec{x}_1) \,
             {\cal O}_X ( 0,\vec{0}) \,
             \chi (t_2,\vec{x}_2) \right) \right| 0 \, \right>
\label{eq:three-pt}
\end{equation}
which has a fermion operator inserted at the spacetime origin between
two external $B$-meson operators, $\chi$.  The times are restricted
to the range $t_1 > 0 > t_2$.  We use the spatially extended
$B$-meson interpolating field
\begin{equation}
\chi( \vec{x} ,t )
  =   \sum_{\vec{r}} f(\vec{r}) \; \overline{q} (t , \vec{x}+\vec{r})
                                \; \gamma_{5} \; b( t , \vec{x})
\label{eq:smear}
\end{equation}
where $f$ is a smearing function chosen~\cite{Draper94a} to project
out the ground state efficiently.  The four-fermion operators, ${\cal
O}_X$ (with $X\in\{L,R,N,S\}$), are defined\footnote{We choose a
standard normalization for which the VSA value for ${\cal O}_{L}$ is
$(8/3)f_B^2 m_B^2$.} as~\cite{Flynn91a}:
\begin{eqnarray}
{\cal O}_{L} & = & \overline{b} \gamma_{\mu} (1-\gamma_5) q
                   \overline{b} \gamma_{\mu} (1-\gamma_5) q \nonumber
\\
{\cal O}_{R} & = & \overline{b} \gamma_{\mu} (1+\gamma_5) q
                   \overline{b} \gamma_{\mu} (1+\gamma_5) q \nonumber
\\
{\cal O}_{N} & = &
             \left( 2 \overline{b} (1-\gamma_5) q \overline{b}
(1+\gamma_5) q
                  + 2 \overline{b} (1+\gamma_5) q \overline{b}
(1-\gamma_5) q
                    \right. \nonumber \\ & & \phantom{00} \left.
                  + \overline{b} \gamma_{\mu} (1-\gamma_5) q
                    \overline{b} \gamma_{\mu} (1+\gamma_5) q
                  + \overline{b} \gamma_{\mu} (1+\gamma_5) q
                    \overline{b} \gamma_{\mu} (1-\gamma_5) q  \right)
                   \nonumber \\
{\cal O}_{S} & = & \overline{b} (1-\gamma_5) q \overline{b}
(1-\gamma_5) q
\label{eq:four-ops}
\end{eqnarray}
The operators ${\cal O}_{R}$ and ${\cal O}_{N}$ are introduced in the
lattice and contribute towards ${\cal O}_{L}$ because of the poor
chiral behavior of Wilson quarks.  The operator ${\cal O}_{S}$ is
introduced in the continuum and contributes because of the matching
of full QCD to the static theory.

With the smeared-sink--local-source (SL) two-point function defined
as
\begin{equation}
C_{2}( t_1 )  =  \sum_{\vec{x}_1}
                 \langle \, 0 \mid  {\cal T}
               ( \chi (t_1 , \vec{x}_1 ) \;
                 \overline{b}(0,\vec{0}) \; \gamma_{4}\gamma_{5}
                 \; q(0 , \vec{0})
               ) \mid 0 \, \rangle
\label{eq:two-point}
\end{equation}
the ``raw'' lattice-static parameters, $B_{X}^{}$, are calculated via
the ratio of three- and two-point functions:
\begin{eqnarray}
B_{X}(t_{1} , t_{2} )
& = & \frac{ C_{3,X}( t_{1} , t_{2} ) }
           { \frac{8}{3}
             C_{2}( t_{1} )
             C_{2}( t_{2} ) }
    \ \stackrel{ \left| t_i \right| \,\gg 1 }
               { -\!\!\!-\!\!\!-\!\!\!\longrightarrow }\
      B_{X}^{}
\label{eq:BnumerDEF}
\end{eqnarray}
The $B_B^{}$ parameter itself can then be determined from the $B_X
\equiv B_{{\cal O}_X}$, extracted from fits of the Monte Carlo data
to the form of Eq.~(\ref{eq:BnumerDEF}), as the linear combination
\begin{eqnarray}
B_{B}^{}
& = &   Z_{B_L} B_{L} + Z_{B_R} B_{R}
      + Z_{B_N} B_{N} + Z_{B_S} B_{S}
\label{eq:fit-combine}
\end{eqnarray}
where the perturbatively-calculated mixing coefficients, $Z_{B_X}$,
are defined in Sec.~\ref{Sec:PerMat}.  Rather than this
``fit-then-combine'' method, our quoted results will be from the
``combine-then-fit'' method:
\begin{eqnarray}
B_{B}(t_{1} , t_{2} )
& = & \sum_{X=L,R,N,S} Z_{B_X} B_{X}(t_{1} , t_{2} )
    \ \stackrel{ \left| t_i \right| \,\gg 1 }
               { -\!\!\!-\!\!\!-\!\!\!\longrightarrow }\
      B_{B}^{}
\label{eq:combine-fit}
\end{eqnarray}
For infinite statistics, the two methods should give identical
results.

We exploit time-reversal symmetry by averaging the correlators over
$t$ and $T-t$, where $T$ is the length of the lattice in the time
direction.  We fix one of the times, $t_{1}$, in
Eqs.~(\ref{eq:BnumerDEF}) and~(\ref{eq:combine-fit}) and vary the
other, $t_{2}$; the result is fitted to a constant.  The fits include
correlations in time, but not in the chiral extrapolation (a choice
forced upon us by our limited statistics).  The entire fitting
procedure is bootstrapped (see, for example, Ref.~\cite{Ukqcd93a}) to
provide robust estimates of the statistical errors.  An estimate of
the systematic error due to the choice of interval is made by
calculating the variance of the results from using all ``reasonable''
time intervals around our favorite one.

A major problem with simulations that include static quarks is that
the signal-to-noise ratio decreases very quickly with
time~\cite{Lepage92a,Duncan92a,Bernard92a}; therefore, the operator
which creates the $B$-meson must project onto the ground state at
very early times --- before the signal is lost in the noise.
Experience with the calculation of the $f_{B}^{}$ decay constant in
the static theory has shown that reliable results can be obtained
only if the $B$-meson operator is smeared with a very accurate ``wave
function,'' which can be obtained from a variational calculation on
the lattice.  We use the same smearing function as was used in our
calculation of $f_{B}^{}$ in the static approximation.  This was
obtained from the variational technique, called {\sc
most}~\cite{Draper94a}, which we have developed for this purpose.

\begin{figure}[t]
\def\filename{LSeffmass.ps}
\begin{center}
\epsfysize=\hsize \leavevmode
\rotate[l]{\epsfbox{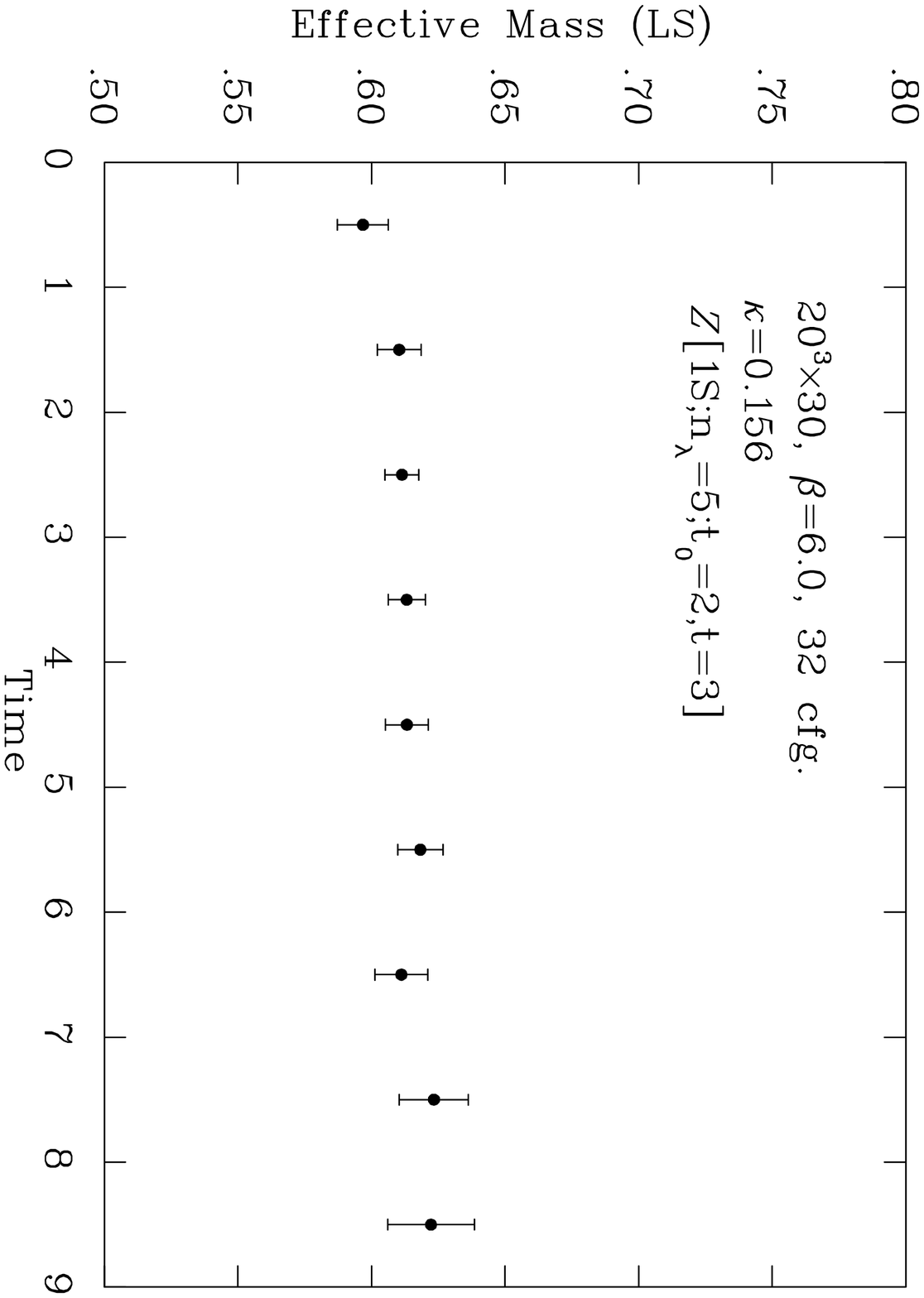}}
\end{center}
\caption{\label{LS_effmass}
Effective mass $m(t+1/2)=\ln C(t)/C(t+1)$ from the LS (local sink,
smeared source) two-point correlation function $C(t)$.  The source
was smeared with an optimal smearing function produced by the {\sc
most}~\protect\cite{Draper94a} algorithm which was designed to
eliminate excited-state contamination.}
\end{figure}

\begin{figure}[t]
\def\filename{SLeffmass.ps}
\begin{center}
\epsfysize=\hsize \leavevmode
\rotate[l]{\epsfbox{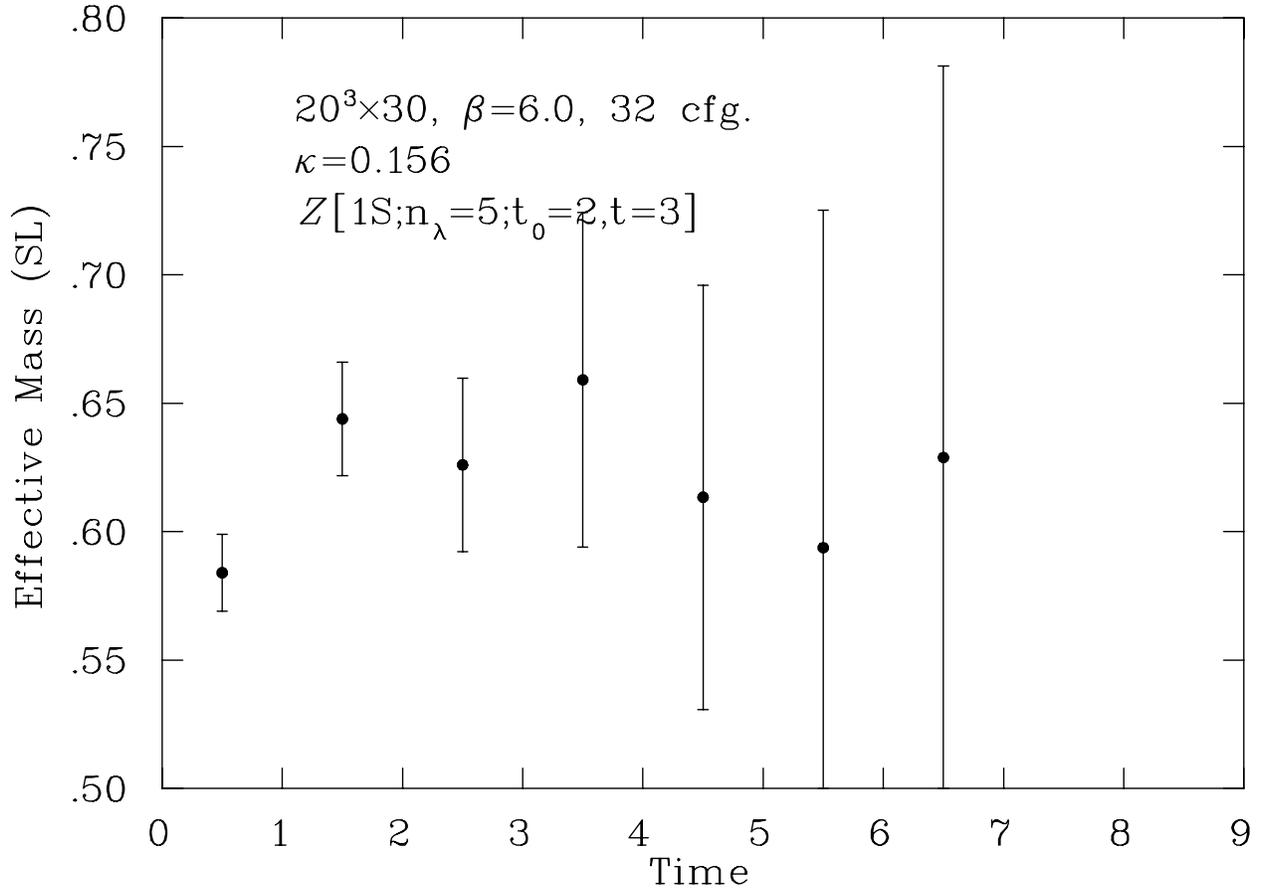}}
\end{center}
\caption{\label{SL_effmass}
Same as for Fig.~\protect\ref{LS_effmass} but for the SL (smeared
sink, local source) two-point correlation function.  The same optimal
smearing function is used to eliminate excited state contamination,
but statistical errors are larger since the source is (necessarily) a
delta function.}
\label{\filename}
\end{figure}

To demonstrate the effectiveness of the smearing function produced by
{\sc most}, we show in Fig.~\ref{LS_effmass} the effective-mass plot
($\ln C_{2}^{LS}(t)/C_{2}^{LS}(t+1)$ versus $t+1/2$) for the
two-point correlation function using a local (delta-function) sink at
time $t$ and an optimally-smeared source at time $0$.  The
effective-mass plot has plateaued at small $t$ (indicating the
absence of significant excited-state contamination) before the
signal-to-noise ratio has degenerated, so that a very precise mass
and amplitude can be obtained by fitting over an early time range.
If, instead, the same smearing function is used at the sink, with a
local (delta-function) source, then it will still effectively remove
excited-state contamination.  Yet, as demonstrated in
Fig.~\ref{SL_effmass}, this fact is obscured by much larger
statistical fluctuations.  (Since the spatial points are summed over
at the sink to project out zero momentum regardless of which smearing
function is used, smearing at the sink provides only marginal
improvement in the signal and increases noise.  In contrast, smearing
at the source greatly enhances the signal and decreases the noise.
For the local source the static quark is restricted to the spatial
origin, and thus the statistics are poorer~\cite{Bernard94a}.)

We note that once an ``optimal'' smearing source has been obtained
from the two-point function using a variational technique, it can be
used directly in other calculations.  The three-point function does
not need to be formulated as a variational problem, although
ground-state dominance should still be monitored using the mass
splitting between the excited and ground states.

\begin{figure}[t]
\def\filename{smear.ps}
\begin{center}
\epsfysize=\hsize \leavevmode
\rotate[l]{\epsfbox{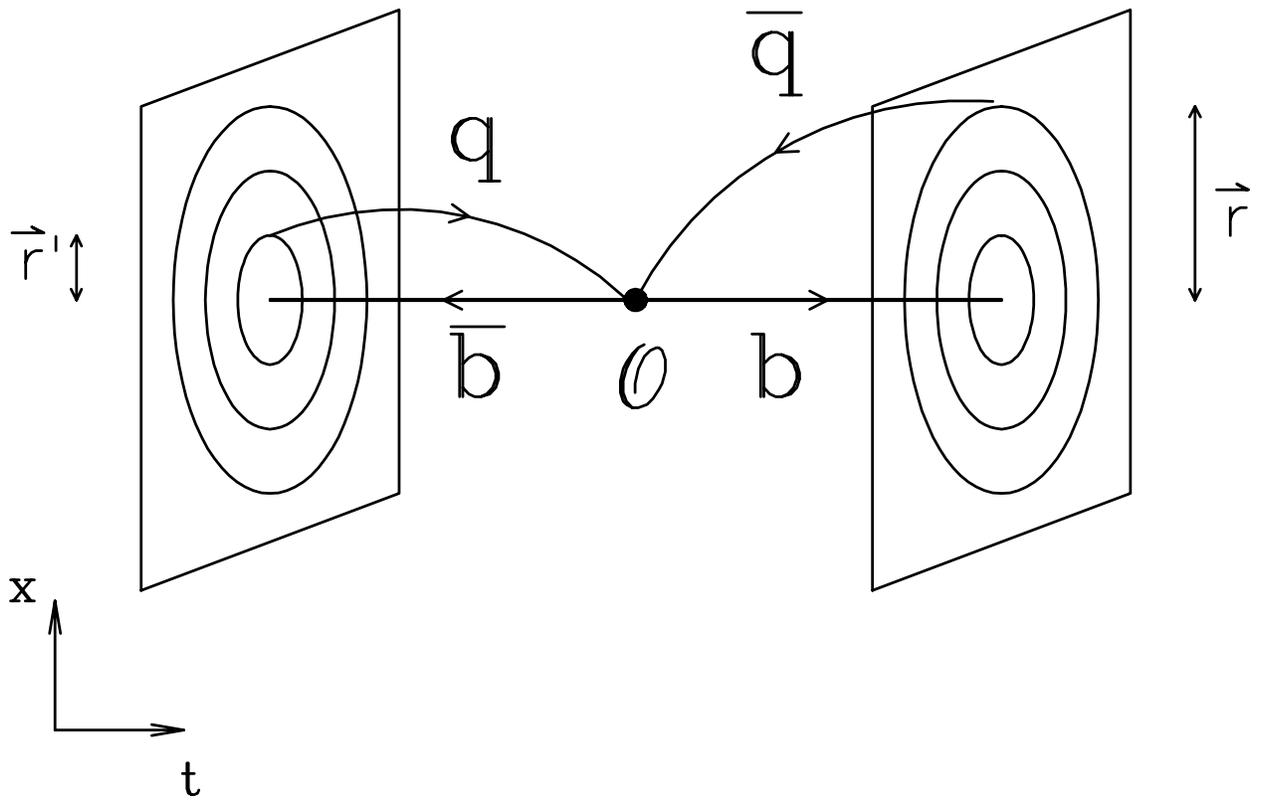}}
\end{center}
\caption{\label{smear}
Schematic diagram of the quark flow for the three-point correlation
function of Eq.~(\protect{\ref{eq:three-pt}}).  The ``targets'' are
intended to represent the smearing of the light quark relative to the
static quark.  The static quarks are restricted to the spatial
origin.}
\end{figure}

The static quark never evolves in space from the origin because the
four-fermion operator is at the spacetime origin.  The $B$-meson
operator is constructed by smearing the light quark relative to the
heavy quark~(Eq.~(\ref{eq:two-point})).  Fig.~\ref{smear} shows a
schematic of the quark flow resulting from the Wick contraction of
Eq.~(\ref{eq:three-pt}).  The resulting two-point correlators are
smeared-sink--local-source (SL) correlators, which are much more
noisy than local-smeared (LS) correlators (as argued above) even
though in the infinite-statistics limit they are equal.  Since
three-point functions are, in general, noisier than two-point
functions, the ``effective-mass'' plots for these are even noisier
than that for the SL two-point function; it would be hopeless to get
a precise result for a static-light-meson matrix element without
using a prohibitively large number of configurations.  But
fortunately, because the $B_{B}^{}$ parameter is a {\sl ratio\/} of
matrix elements (Eq.~(\ref{eq:BnumerDEF})), the noise is reduced due
to the cancelation of correlated fluctuations between the numerator
and denominator.

It has been argued that the product $B_{B}^{} f_{B}^{2}$ and
perturbative corrections to it should be calculated directly since
it, rather than $B_{B}^{}$, is the phenomenologically-important
quantity.  But there are several compelling reasons for calculating
$f_{B}^{}$ and $B_{B}^{}$ separately.  Firstly, although the
calculation of $B_{B}^{}$, as for $B_{B}^{} f_{B}^{2}$, is
intrinsically more involved than is that of $f_{B}^{}$ (both
analytically, in the determination of perturbative corrections, and
computationally), the numerical value of $B_{B}^{}$ is more stable
than is the value of either $B_{B}^{} f_{B}^{2}$ or $f_{B}^{}$.
Certainly, $f_{B}^{}$ is a very important physical quantity in its
own right; it should be and is calculated separately.  For this, one
need only calculate a two-point function.  However, the statistical
fluctuations for $f_{B}^{}$ are quite large; without the use of a
reliable smearing function obtained variationally, excited-state
contamination can be substantial and can mislead interpretation.
(This may explain the scatter in the world summary of lattice
calculations of $f_{B}^{}$~\cite{Bernard94a,Allton96a}.)  Also, since
its lattice-spacing dependence is rather large, especially when using
the static approximation, its continuum extrapolation is delicate and
prone to large systematic errors.  Much computing effort is required
to evaluate this simple quantity.  However, $B_{B}^{}$ (or $B_{B}^{}
f_{B}^{2}$) requires the calculation of a three-point, in addition to
a two-point, correlation function.  Since it is more involved, it is
usually determined in a secondary calculation after the primary
calculation of $f_{B}^{}$ and so fewer groups are likely to calculate
it.  Yet, as borne out by our data, since $B_{B}^{}$ can be extracted
from a ratio of three- to two-point functions which are strongly
correlated, a quite precise value can be obtained, with an optimal
choice of smearing function, from relatively few configurations.  The
calculational overhead (both computational and analytical) is large
compared to the computational expense.  Thus a handful of groups can
fix precisely the value of $B_{B}^{}$ once and for all, leaving for
the wider community the task of applying improvements in algorithms
and computers to the simpler $f_{B}^{}$.  In the future, $B_{B}^{}$
(in contrast to $f_{B}^{}$ and $B_{B}^{} f_{B}^{2}$) need not be
recalculated with every generation of improvements.

Secondly, just as the numerical value of $B_{B}^{}$ is stable because
of cancelations of correlated fluctuations in numerator and
denominator, we argue that so too are its perturbative corrections
when linearized as is demonstrated in Sections~\ref{Sec:PerMat}
and~\ref{Sec:SysErr}.  The perturbatively-calculated coefficients for
$B_{B}^{}$ are likely more reliable than those for the product
$B_{B}^{} f_{B}^{2}$.  Likewise, these are less likely to need
updating with the next generation of improvements in analytic
methods.

Thirdly, it seems as though VSA is a surprisingly good approximation
for the $B_{B}^{}$ parameter.  This is an important qualitative
statement, of use to model builders, which should not be obscured by
poor-statistics attempts to calculate the product $B_{B}^{}
f_{B}^{2}$.

\section{Numerical results}
\label{Sec:NumRes}

The simulations were carried out on a $20^{3} \times 30$ lattice,
calculated on 32 gauge configurations, at $\beta = 6.0$.  (This
number of configurations is more than adequate for a precise estimate
of the $B_{B}^{}$ parameter with small statistical error since an
efficient smearing function is used.  The use of an {\it ad hoc\/}
smearing function would have required an order of magnitude more
configurations.)  The simulations were quenched; the gauge
configurations were generated using the standard Wilson pure-glue
action.  The gauge configurations were fixed into Coulomb gauge.  (An
ultra-conservative gauge-fixing convergence criterion was used such
that $\vec{\nabla}\!\cdot\!\vec{A}$ was decreased to less than
$10^{-9}$ its unfixed value.)  The gauge-fixing was done only to
choose smearing functions, but since these cancel in ratios of
correlation functions all results are gauge-invariant (in the
infinite-statistics limit).  Wilson light-quark propagators, with
hopping-parameter values, $\kappa=0.152$, $0.154$, $0.155$, and
$0.156$, were used in our analysis.  The value of kappa-critical used
was $0.157$ and the value of kappa-strange was
$0.155$~\cite{Bhattacharya95a}.

\begin{figure}[t]
\def\filename{OL_bparam.ps}
\begin{center}
\epsfysize=\hsize \leavevmode
\rotate[l]{\epsfbox{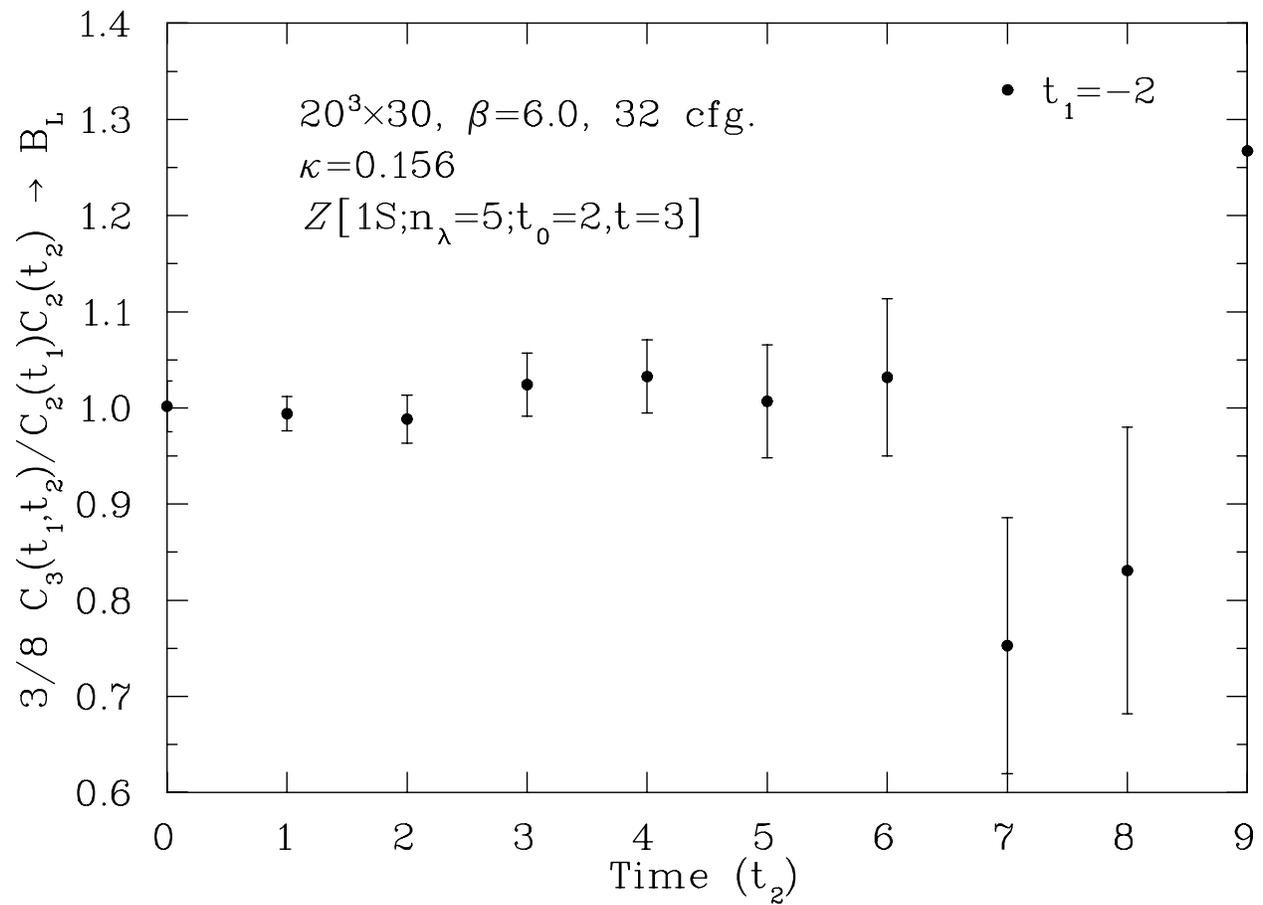}}
\end{center}
\caption{\label{OL_bparam}
Raw $B$ parameter for the ${\cal O}_{L}$ operator from
Eq.~(\protect\ref{eq:BnumerDEF}). }
\end{figure}

Fig.~\ref{OL_bparam} shows, for the operator ${\cal O}_{L}$, the
ratio of the three- and two-point correlation functions
$B_{L}(t_{1},t_{2})$, Eq.~(\ref{eq:BnumerDEF}), which asymptotically
equals $B_{L}$ for large Euclidean times.  (In the figure
$B_{L}(t_{1},t_{2})$ is graphed as a function of $t_{2}$ with
$t_{1}=-2$ held fixed.)  In fact, ``large times'' are remarkably
small ($>\approx 2$) because of the effectiveness of the smearing
function in efficiently eliminating excited-state contamination, a
fact supported by Fig.~\ref{LS_effmass}.

\begin{figure}[t]
\def\filename{OL_fit.ps}
\begin{center}
\epsfysize=\hsize \leavevmode
\rotate[l]{\epsfbox{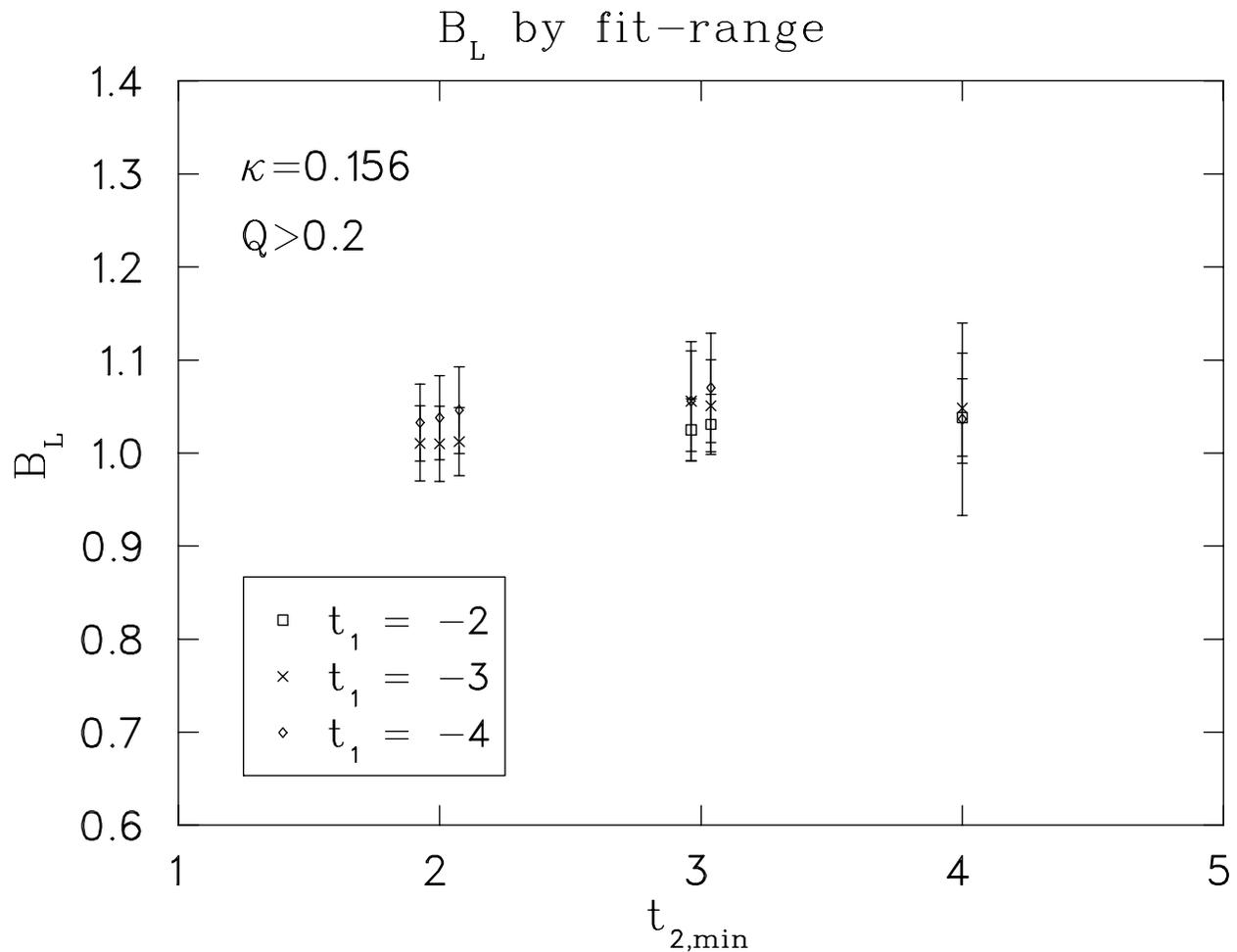}}
\end{center}
\caption{\label{OL_fit}
The dependence on the fitted raw $B_{L}$ parameter on the choice of
$t_{1}$, the (fixed) time position of one interpolating field, and on
the fit range $t_{2,{\rm min}}^{} - t_{2,{\rm max}}^{}$ of the other.
Clustered points have different $t_{2,{\rm max}}$.  All fits take
into account correlations in $t_{2}$, and are not displayed if the
naive quality of fit $Q$ does not exceed 0.2. }
\end{figure}

As with any lattice calculation of correlation functions, there is
freedom in the choice of fit range and a balance needs to be struck
between fitting over too-early times, for which systematic errors due
to excited-state contamination may be non-negligible, and over
too-late times, for which statistical errors will be unnecessarily
large.  In Fig.~\ref{OL_fit} we display a $t_{\rm min}$--plot: the
values for the fits of the raw $B_{L}$ value (at $\kappa=0.156$)
plotted for several choices of fit range.  (All of our fits take into
account the correlations in Euclidean time using the full-covariance
matrix.  For our central fit range, the values of the fits differ
little whether or not the correlations are included.)  The flatness
of the plateau in Fig.~\ref{OL_bparam} reflects the insensitivity of
the fitted value to the choice of fit range.  For this and other
plots we choose as our central values $t_{1}=-2$ and $3 \le t_{2} \le
6$, a moderately-aggressive choice which has good $\chi^{2}/{\rm
dof}$ ($0.83/3$, $0.59/3$, $0.41/3$, $0.68/3$ for $\kappa=0.152$,
$0.154$, $0.155$, $0.156$, respectively), small statistical errors,
and fit-range systematic errors which are smaller than, but
comparable to, the statistical errors.  The fit-range systematic
errors are determined from the standard deviation of all ``reasonable
choices.''

\begin{figure}[t]
\def\filename{OR_bparam.ps}
\begin{center}
\epsfysize=\hsize \leavevmode
\rotate[l]{\epsfbox{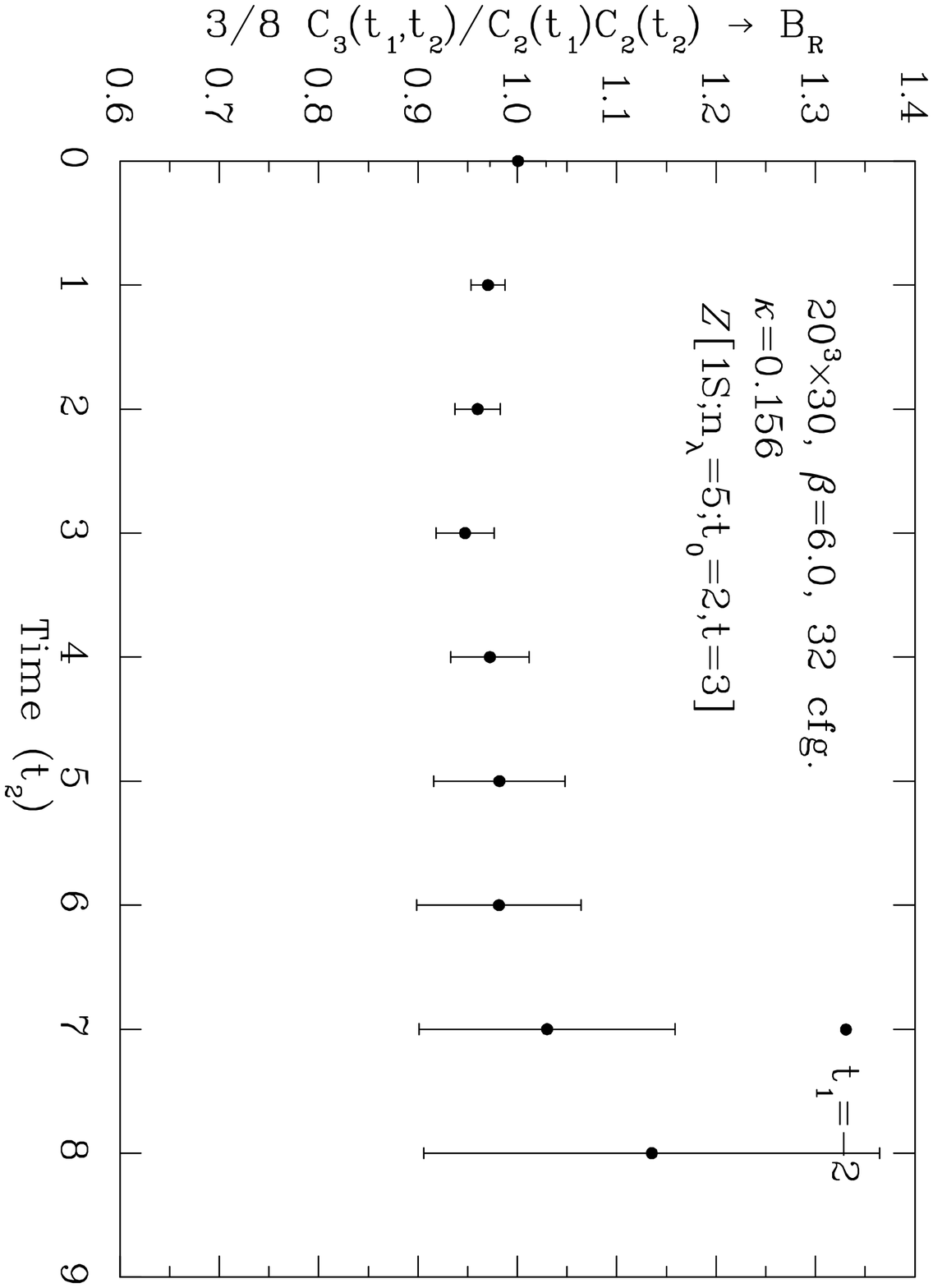}}
\end{center}
\caption{\label{OR_bparam}
Same as for Fig.~\protect\ref{OL_bparam} but for the ${\cal O}_{R}$
operator.}
\end{figure}

\begin{figure}[t]
\def\filename{ON_bparam.ps}
\begin{center}
\epsfysize=\hsize \leavevmode
\rotate[l]{\epsfbox{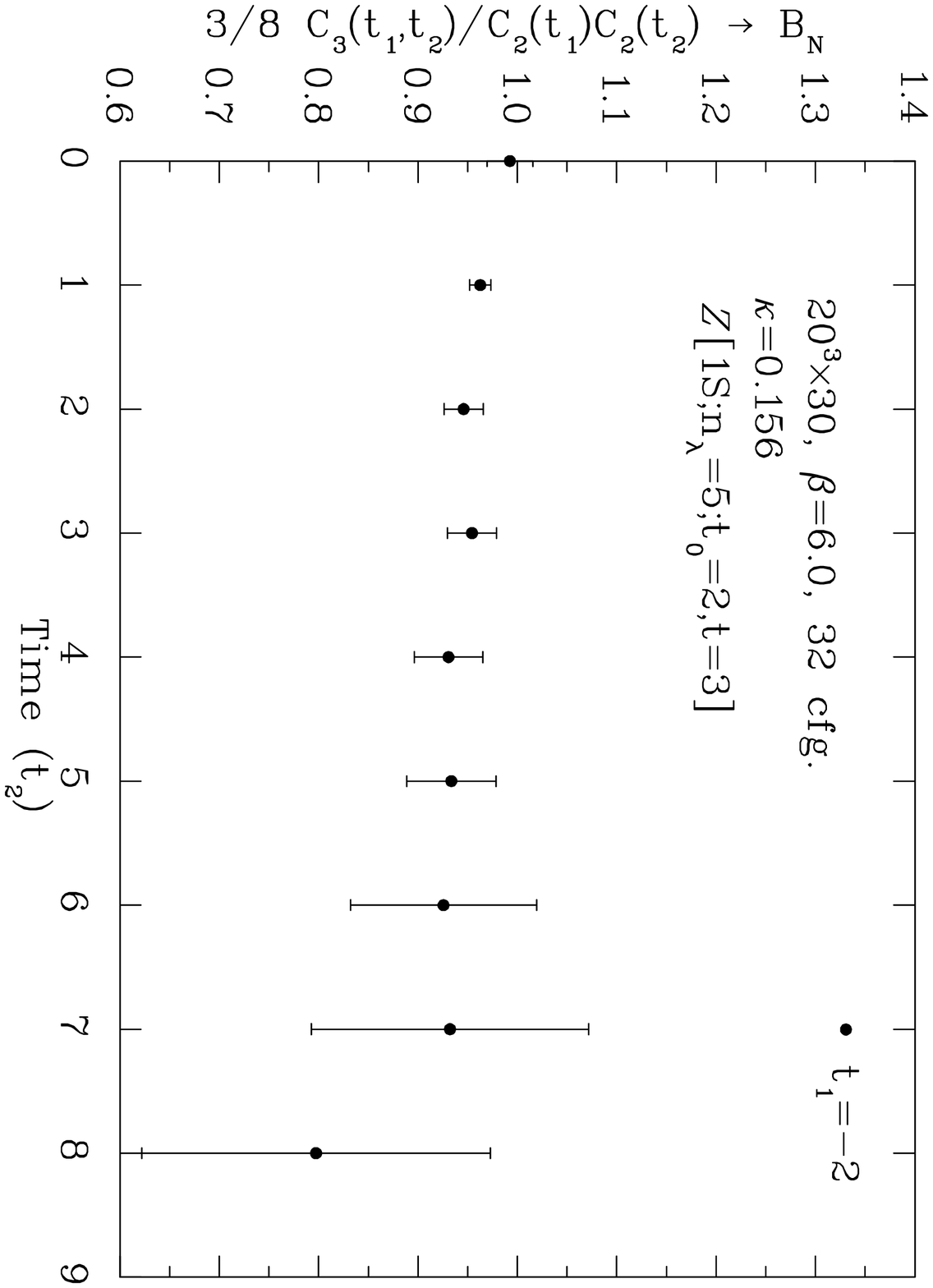}}
\end{center}
\caption{\label{ON_bparam}
Same as for Fig.~\protect\ref{OL_bparam} but for the ${\cal O}_{N}$
operator.}
\end{figure}

\begin{figure}[t]
\def\filename{OS_bparam.ps}
\begin{center}
\epsfysize=\hsize \leavevmode
\rotate[l]{\epsfbox{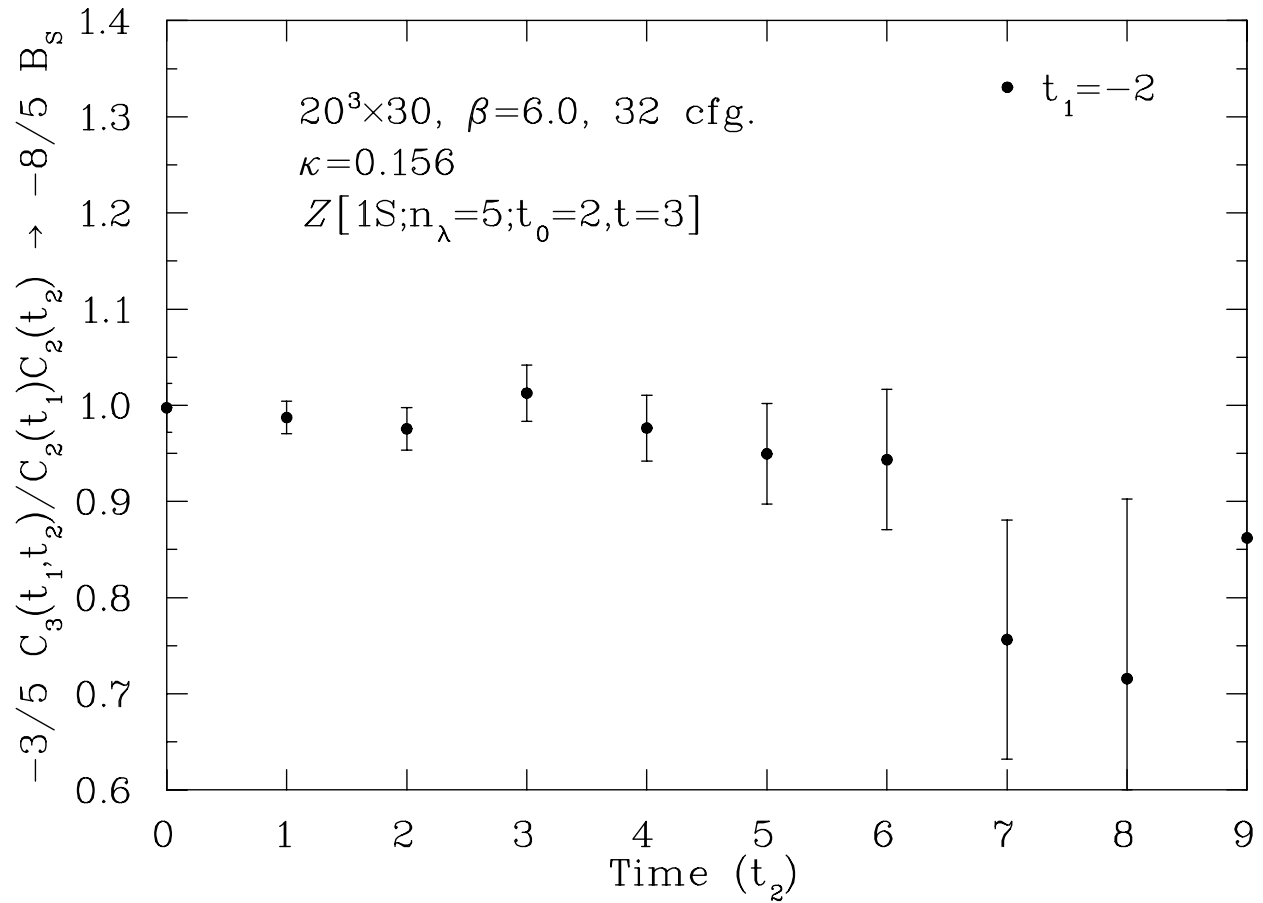}}
\end{center}
\caption{\label{OS_bparam}
Same as for Fig.~\protect\ref{OL_bparam} but for the ${\cal O}_{S}$
operator.  Note the normalization as explained in
Table~\protect\ref{tb:Bparam}. }
\end{figure}

\begin{figure}[t]
\def\filename{OB_bparam.ps}
\begin{center}
\epsfysize=\hsize \leavevmode
\rotate[l]{\epsfbox{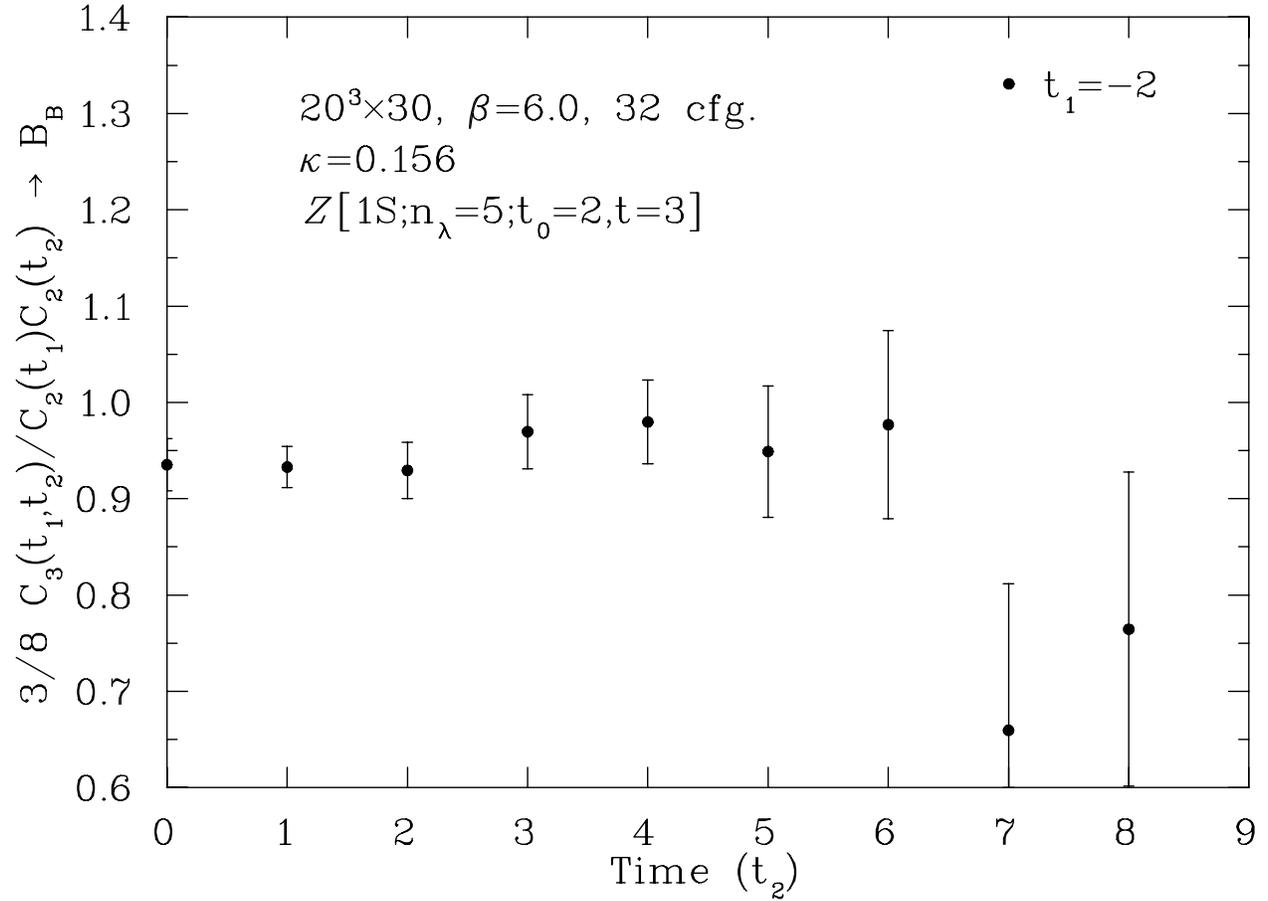}}
\end{center}
\caption{\label{OB_bparam}
The ratio of (the linear combination of) three-point functions to
two-point functions which approaches $B_{B}^{}$ for large Euclidean
times.}
\end{figure}

\begin{figure}[t]
\def\filename{BB_fit.ps}
\begin{center}
\epsfysize=\hsize \leavevmode
\rotate[l]{\epsfbox{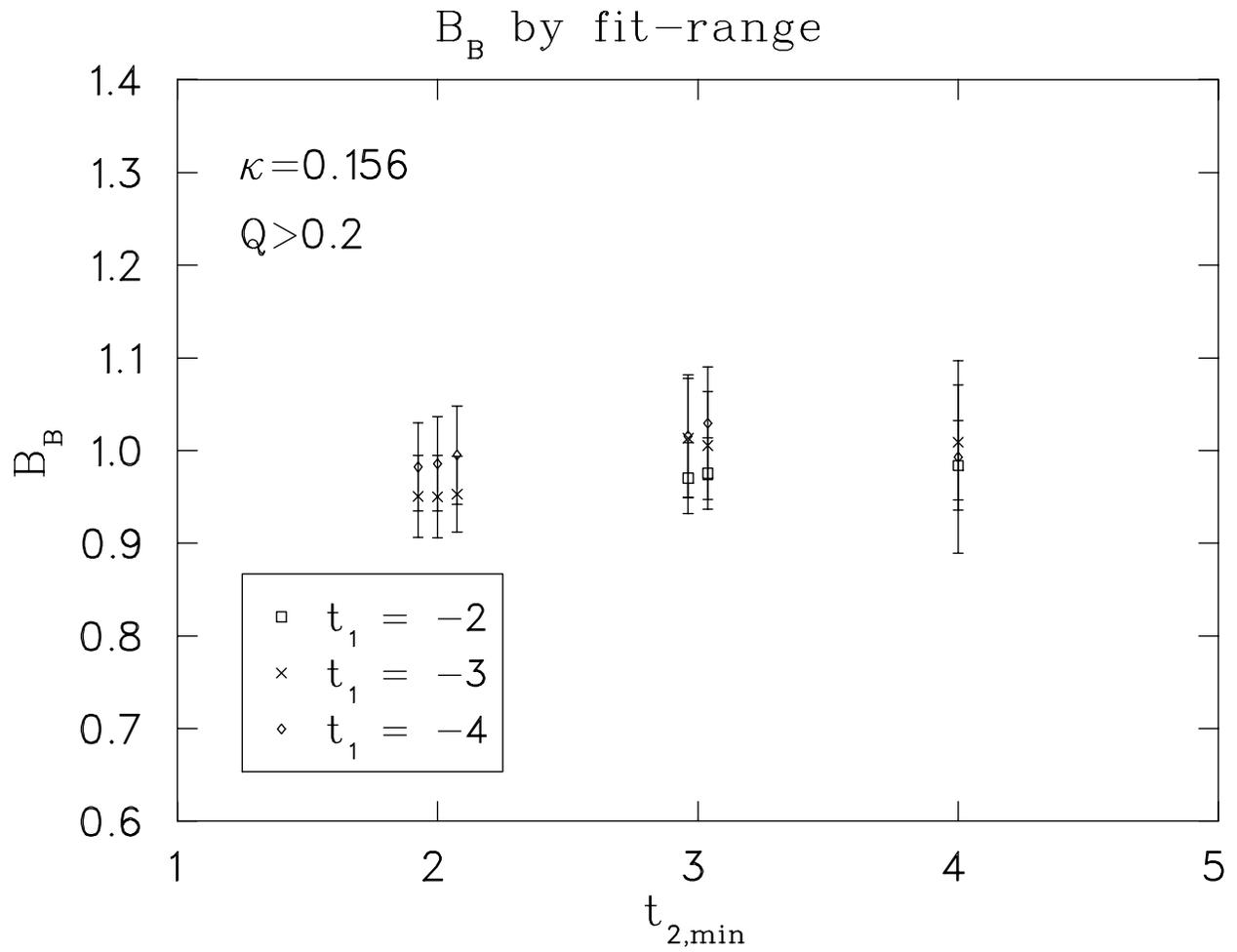}}
\end{center}
\caption{\label{BB_fit}
Same as for Fig.~\protect\ref{OL_fit} but for the $B_{B}^{}$
parameter itself.}
\end{figure}

Figs.~\ref{OR_bparam},~\ref{ON_bparam}, and~\ref{OS_bparam} show
similar plots for the raw lattice values for $B_{R}$, $B_{N}$ and
$B_{S}$, respectively.  Fig.~\ref{OB_bparam} shows the ratio of
correlation functions defined in Eq.~(\ref{eq:combine-fit}) from
which the desired $B_B^{}$ parameter is extracted.  Again, the plot
plateaus early with small statistical errors.  Fig.~\ref{BB_fit}
shows that, again, the value is insensitive to the choice of fit
region.  (For our central choice of fit range, the $\chi^2/{\rm dof}$
are $0.74/3$, $0.57/3$, $0.43/3$, $0.67/3$ for $\kappa=0.152$,
$0.154$, $0.155$, $0.156$, respectively.)  We could also calculate
the final $B_{B}^{}$ parameter from the appropriate linear
combination of the four fitted raw values $B_{L}$, $B_{R}$, $B_{N}$
and $B_{S}$, as in Eq.~(\ref{eq:fit-combine}).  The $\chi^{2}/{\rm
dof}$ are good for $B_{R}$ (0.71, 0.67, 0.55, 0.33) and $B_{N}$
(0.60, 0.42, 0.36, 0.40).  The worst $\chi^{2}/{\rm dof}$ are for
$B_{S}$ (1.38, 1.30, 1.14, 0.85).  The two procedures,
combine-then-fit versus fit-then-combine, could give different
answers in principle (for finite statistics), but in practice we see
little difference.

%
%
\begin{table}[t]
\begin{center}
\begin{tabular}{lccccc}
& $\kappa=0.152$ & $\kappa=0.154$ & $\kappa=0.155$
                 & $\kappa=0.156$ & $\kappa_c=0.157$    \\
\hline
$\phantom{-\frac{8}{5}} B_{L}$
     & $1.01^{+2}_{-2}{(1)}$ & $1.02^{+2}_{-2}{(1)}$
     & $1.02^{+3}_{-2}{(1)}$ & $1.03^{+3}_{-3}{(2)}$
     & $1.03^{+3}_{-3}{(2)}$ \\
$\phantom{-\frac{8}{5}} B_{R}$
     & $0.96^{+1}_{-1}{(1)}$ & $0.96^{+1}_{-2}{(2)}$
     & $0.95^{+2}_{-2}{(2)}$ & $0.95^{+2}_{-3}{(2)}$
     & $0.95^{+2}_{-3}{(2)}$ \\
$\phantom{-\frac{8}{5}} B_{N}$
     & $0.97^{+2}_{-2}{(3)}$ & $0.96^{+2}_{-2}{(4)}$
     & $0.96^{+2}_{-2}{(4)}$ & $0.96^{+3}_{-2}{(4)}$
     & $0.95^{+3}_{-2}{(5)}$ \\
$         -\frac{8}{5}  B_{S}$
     & $1.00^{+2}_{-1}{(2)}$ & $1.00^{+2}_{-2}{(2)}$
     & $1.00^{+2}_{-2}{(3)}$ & $1.01^{+3}_{-3}{(3)}$
     & $1.01^{+3}_{-3}{(3)}$ \\
\hline
$\phantom{-\frac{8}{5}} B_{B}(m_b^\star)$
     & $0.95^{+2}_{-2}{(1)}$ & $0.96^{+3}_{-2}{(2)}$
     & $0.96^{+3}_{-3}{(2)}$ & $0.98^{+4}_{-4}{(2)}$
     & $0.98^{+4}_{-4}{(3)}$
\end{tabular}
\end{center}
\caption{\label{tb:Bparam}
The raw lattice $B$ parameters for the operators ${\cal O}_{L}$,
${\cal O}_{R}$, ${\cal O}_{N}$, and ${\cal O}_{S}$ which appear in
the
lattice-continuum matching, and the linear combination $B_{{\cal
O}_{L}^{\rm full}} \equiv B_{B}^{}$ as a function of $\kappa$ and
extrapolated to $\kappa_{c}$.  The first errors are statistical
(bootstrap) and the second are systematic due to choice of fit range.
Note that ${\cal O}_{S}$ has a VSA value different from that of
${\cal
O}_{L}$; with our normalization for the raw $B$ parameters (a common
denominator equal to the VSA value of ${\cal O}_{L}$)
$-\frac{8}{5}B_{S}$ would identically equal 1 if VSA were exact.}
\end{table}
As shown in Table~\ref{tb:Bparam}, each raw lattice $B$ parameter is
close to 1.0 with small statistical errors, so our final
value\footnote{\label{f:m_b_star} $B_B$ is evaluated at $m_b^\star$,
which is the scale at which the running mass is $m(m_b^\star) \!=\!
m^{}_{b_{pole}} \!=\!
4.72\,$GeV~\cite{Duncan94c,Dominguez92,PDG96}.}  for
$B_{B}^{}(m_b^\star)$ is also close to 1.0, the VSA value, with
similarly small statistical errors.

\section{Perturbative matching}
\label{Sec:PerMat}

To calculate the continuum value of the $B_{B}^{}$ parameter, our
``raw'' lattice results, listed in Table~\ref{tb:Bparam}, must be
multiplied by a lattice-to-continuum perturbative matching
coefficient.  After we finished the first analysis of our
data~\cite{Draper96a}, we found that our value for $B_B^{}$ was
approximately $30\%$ higher than the result of a similar simulation
by the UKQCD collaboration~\cite{Ukqcd96a}.  We suspected that this
difference was due to more than just the difference in the actions.
This motivated us to do a very careful study of the perturbative
matching, using the results in the literature, so that we obtained
the ``best value'' of $B_B^{}$ using the information available to us.
(This is discussed further in
Secs.~\ref{Sec:SysErr},~\ref{Sec:Compare}, and~\ref{Sec:Con}.) We
also studied the systematic errors in the perturbative matching to
find the reason for the disagreement between UKQCD's result and ours.

For convenience, we shall refer to the $\Delta B = 2$ effective
Hamiltonian, obtained from the standard model by integrating out the
top quark and the heavy vector gauge bosons, as the ``full'' theory
although this is also an effective field theory.  The perturbative
matching is broken into two stages: full QCD to the continuum-static
theory and the continuum-static theory to the lattice-static theory.
For the matching of full QCD to the continuum-static theory, the
relevant perturbative results have been calculated to do a
next-to-leading-order analysis of the $\log(\mu/m_b)$ terms.  The use
of renormalization-group-improved perturbation theory reduces the
renormalization-scheme dependence and the effects of the different
ways of defining $\gamma_{5}$ in dimensional
regularization~\cite{Buras90b}.  Two scales are necessary for the
perturbative matching: the scale, $\mu_b \! = \! {\it
O}\!\left(\!m_b\!\right)$, of the matching to the full theory (we
choose $\mu_b \!=\! m_b^\star$, where $m_b^\star$ is defined as
mentioned earlier in footnote~\ref{f:m_b_star}) and the scale, $\mu$,
of the matching to the lattice theory (we choose $\mu \!=\!
q^\star$, which is determined from the Lepage-Mackenzie scale
formulation~\cite{Lepage93} as discussed later in this section).
Also, as emphasized by Ciuchini {\it et al.}~\cite{Ciuchini95a}, it
is important to check the stability of the perturbative coefficient
at next-to-leading order as the renormalization scale is changed.

We choose, as do others~\cite{Ukqcd96a,Ciuchini96b,Buchalla97a}, to
evaluate the full-theory operator, ${\cal O}^f$, at $\mu_b$:
\begin{equation}
\left< {\cal O}^{f}\!\left(\!\mu_b\!\right) \right>
  =  \sum_{i=L,S}
     C_{\phantom{f}i}^{fc}\!\left(\!\mu_b; \mu\!\right)
     \left< {\cal O}_{i}^{c}\!\left(\!\mu\!\right) \right>
\label{eq:ope_like}
\end{equation}
where terms of order $1/m$ have been dropped.  We use a
double-argument notation similar to Ref.~\cite{Buchalla97a} to
emphasize that this matching of the continuum-static theory to the
full theory involves two theories ($f$ and $c$) and two scales
($\mu_b$ and $\mu$).  $C_{\phantom{f}i}^{fc}\!\left(\!\mu_b;
\mu\!\right)$ includes a running of the scale in the continuum-static
theory which can be written explicitly due to the form of the
solution to the renormalization group equation (RGE) for the
coefficients (see, for example,~\cite{Buras92a}).
\begin{eqnarray}
C^{fc}_{\phantom{f}i}\!\left(\!\mu_b; \mu\!\right)
& = & C^{fc}_{\phantom{f}j} \!\left(\!\mu_b; \mu_b\!\right) \;
     \left ( {\cal T}_g
      \exp \!\left\{-\int^{g^c\!\left(\!\mu_b\!\right)
                        }_{g^c\!\left(\!\mu\!\right)}
                     d\overline{g}
                     \frac{ \hat{\gamma}^c
                            \!\left(\!\overline{g}\!\right) }
                          {      \beta^c
                            \!\left(\!\overline{g}\!\right) }
                     \right\} \right )_{ji}
      \nonumber \\
& \equiv & C^{fc}_{\phantom{f}j} \!\left(\!\mu_b; \mu_b\!\right)
      \left( \widehat{U}^T \right)^{c}_{ji}
    \!\left(\!\mu_b,\mu\!\right)
      \label{eq:c_mub_mu}
\end{eqnarray}
Since we focus on the transformation of the operators, we treat the
coefficients, $C$, as a row vector and transpose ($T$) the matrix $U$
to be consistent with the common notation for
$U$~\cite{Neubert94,Buchalla95b} which treats the coefficients as a
column vector: $(C^T)^{cf}_{i} \! ( \mu; \mu_b ) =
\widehat{U}^{c}_{ij} \! (\mu, \mu_b ) \; (C^T)^{cf}_{j} \! ( \mu_b;
\mu_b )$ for which
\begin{eqnarray}
\widehat{U}^{c} \!\left(\!\mu,\mu_b\!\right)
& = & {\cal T}_g \exp \!\left( \int^{g^c\!\left(\!\mu\!\right)
                                  }_{g^c\!\left(\!\mu_b\!\right)}
                      d\overline{g}
                      \frac{ \hat{\gamma}^{cT}\!\left( \!\overline{g}
                                                       \!\right) }
                           { \beta^c\!\left( \!\overline{g}
                                             \!\right) }
                      \right)
\end{eqnarray}
The superscript-label $c$ indicates that the variables are for the
continuum-static theory in which some degrees of freedom have been
removed.  Notice that the continuum-static scale-evolution matrix
scales only the static-theory argument of the coefficient.  Thus,
Eq.~(\ref{eq:ope_like}) becomes
\begin{equation}
\left< {\cal O}^{f}\!\left(\!\mu_b\!\right)  \right>
  =  \sum_{i,j=L,S}
     C_{\phantom{f}i}^{fc}\!\left(\!\mu_b; \mu_b\!\right)
     \; \left( \widehat{U}^T\right)_{ij}^{c}
     \! \left(\!\mu_b,\mu\!\right)
     \; \left< {\cal O}_{j}^{c}\!\left(\!\mu\!\right) \right>
\label{eq:ope_like_2}
\end{equation}
which is read, right-to-left, as ``The static theory operator is
scaled from $\mu$ to $\mu_b$ where it is matched to the full
theory.''

An alternative, not used here, is to evaluate the full theory
operator at the same scale $\mu$ as is the static-theory operator, so
that
\begin{eqnarray}
\left< {\cal O}^{f}\!\left(\!\mu\!\right) \right>
& = & \sum_{i=L,S}
      C_{\phantom{f}i}^{fc}\!\left(\!\mu; \mu\!\right)
   \; \left< {\cal O}_{i}^{c}\!\left(\!\mu\!\right)  \right>
      \nonumber \\
& = & \sum_{i,j=L,S}
      \left(\widehat{U}^T\right)^{f} \!\left(\!\mu,\mu_b\!\right)
   \; C_{\phantom{f}i}^{fc}\!\left(\!\mu_b; \mu_b\!\right)
   \; \left(\widehat{U}^T\right)_{ij}^{c}
\!\left(\!\mu_b,\mu\!\right)
   \; \left< {\cal O}_{j}^{c}\!\left(\!\mu\!\right) \right>
      \label{eq:ope}
\end{eqnarray}
(The generalization to multiple full-theory operators would include
full-theory subscripts on ${\cal O}^{f}$, $(\widehat{U}^T)^{f}$, and
$C_{\phantom{f}i}^{fc}$.)  Eq.~(\ref{eq:ope}) reads, right-to-left,
``The continuum-static theory operator is scaled in the static theory
from $\mu$ to $\mu_b$ where it is matched the full theory and then
scaled back from $\mu_b$ to $\mu$ in the full theory.''  If $U$ is
treated to lowest order, summing neither the leading nor sub-leading
order logarithms, then this reduces to the approach used by Flynn
{\it et al.}~\cite{Flynn91a} who do not use the RG\@.  The
full-theory anomalous dimension appears there since this approach
includes running the scale in the full theory.

Returning to Eqs.~(\ref{eq:ope_like}) and~(\ref{eq:ope_like_2}),
matching in the continuum (with $\mu_b\!=\!m_b^\star$ and
$\mu\!=\!q^\star$) gives
\begin{equation}
\left< {\cal O}^{f}(m_b^\star) \right>
  =   C_L^{fc}(m_b^\star;q^\star)\left<{\cal O}^c_L(q^\star)\right>
    + C_S^{fc}(m_b^\star;q^\star)\left<{\cal O}^c_S(q^\star)\right>
\label{eq:LplusS}
\end{equation}
We use the solution of the RG equation for a matrix of operators
which is discussed by Ciuchini {\it et al.}~\cite{Ciuchini96b} and
Buchalla~\cite{Buchalla97a} in more detail.
\begin{mathletters}
\label{eq:coeffs}
\begin{eqnarray}
C_L^{fc}\!\left(m_b^\star; q^\star\right)
& = & C_L^{fc}\!\left(m_b^\star; m_b^\star\right)
      \left( \frac{ \alpha_s^c\!\left(\!m_b^\star\!\right) }
                  { \alpha_s^c\!\left(\! q^\star \!\right) }
             \right)^{p_{0,L}^c}
      \left\{ 1 + \frac{ \alpha_s^c\!\left(\!m_b^\star\!\right)
                       - \alpha_s^c\!\left(\! q^\star \!\right)
                       }{ 4\pi }
                  p_{1,L}^c \right\}
      \nonumber \\ & &
    + C_S^{fc}\!\left(m_b^\star; m_b^\star\right)
      \frac{ \gamma_{0_{S,L}}^c }
           { \gamma_{0_{L,L}}^c - \gamma_{0_{S,S}}^c }
          \left( \left( \frac{\alpha_s^c\!\left(\!m_b^\star\!\right)}
                             {\alpha_s^c\!\left(\! q^\star \!\right)}
                        \right)^{p_{0,L}^c}
               - \left( \frac{\alpha_s^c\!\left(\!m_b^\star\!\right)}
                             {\alpha_s^c\!\left(\! q^\star \!\right)}
                        \right)^{p_{0,S}^c}
                 \right)
      \label{eq:LSmix} \\
C_S^{fc}\!\left(m_b^\star; q^\star\right)
& = & C_S^{fc}\!\left(m_b^\star; m_b^\star\right)
      \left( \frac{ \alpha_s^c\!\left(\!m_b^\star\!\right) }
                  { \alpha_s^c\!\left(\! q^\star \!\right) }
             \right)^{p_{0,S}^c}
      \label{eq:SS}
\end{eqnarray}
\end{mathletters}
with
\begin{equation}
p_{0,i} = \left( \gamma_{0_{i,i}} / \left( 2 b_0 \right) \right)
\ \ {\rm and} \ \
p_{1,i} = \left[ p_{0,i}
                 \left( \gamma_{1_{i,i}} / \gamma_{0_{i,i}}
                      - b_1 / b_0 \right) \right]
\label{eq:the_ps}
\end{equation}
In Table~\ref{t:the_ps} we list the values of the anomalous
dimensions of the various operators required in this calculation (all
calculated using the naive dimensional regularization scheme).  The
coefficients from the first and second terms of the $\beta$-function
are defined as
\begin{eqnarray}
\beta_0  \equiv  \frac{ b_0 }{ 4 \pi }
         =  \frac{ 11 - \frac{2}{3} n_f }{ 4 \pi }
& \: , \; &
\beta_1  \equiv  \frac{ b_1 }{ 16 \pi^2 }
         =  \frac{ 102 - \frac{38}{3} n_f }{ 16 \pi^2 }
\label{eq:betas}
\end{eqnarray}
To obtain the leading-log expressions from the explicit solutions of
the renormalization group equations that we quote, all quantities
with a subscript $1$ are omitted.  In Eq.~(\ref{eq:SS}) the
higher-order terms of $U^c$ have been dropped when multiplied by
$C_S^{fc}$ because $C_S^{fc}$ is of order $\alpha_s$.  We found that
the inclusion of the $C_S^{fc} (U^T)_{S,L}^c$ term in our analysis
was less than 0.05\% of the $C_L^{fc} (U^T)_{L,L}^c$ term; this is
smaller than the few percent effect which was quoted in
Refs.~\cite{Ciuchini96b,Buchalla97a}.  Our ratio of the coupling at
$\mu$ to that at $\mu_b$ was close to 1 because the automatic
scale-setting procedure selected a scale $q^\star$ which was close to
$m_b^\star$.  As the difference between the scales $\mu$ and $\mu_b$
gets bigger, $(U^T)_{S,L}^c$, which includes the leading off-diagonal
terms in the anomalous dimension matrix, gets larger.
\begin{table}
\begin{center}
\begin{tabular}{lccc}
     & their    & our & \\
Ref. & notation & notation & value \\ \hline
\begin{tabular}{l}
  Ciuchini {\it et al.}~\cite{Ciuchini96b},
  Buchalla~\cite{Buchalla97a}
  \phantom{$\gamma_{11}^{(0)}$} \\
  Gim\'enez\cite{Gimenez93a}
  \phantom{$\left. 4\gamma_{+}^{(1)} \right|_{\overline{MS}}$}
\end{tabular} &
\begin{tabular}{c}
  $\gamma_{11}^{(0)}$ \\
  $\left. 4\gamma_{+}^{(1)} \right|_{\overline{MS}}$
\end{tabular}
  & $\gamma_{0_{L,L}}^c$ & $-8$ \\ \hline
\begin{tabular}{l}
  Ciuchini {\it et al.}~\cite{Ciuchini96b},
Buchalla~\cite{Buchalla97a}
  \phantom{$\gamma_{11}^{(1)}$} \\
  Gim\'enez\cite{Gimenez93a}
  \phantom{$\left. \gamma_{{\cal O}_{\Delta
B=2}}^{(2)}\right|_{NDR}$}
\end{tabular} &
\begin{tabular}{c}
  $\gamma_{11}^{(1)}$ \\
  $16\left. \gamma_{+}^{(2)}\right|_{\overline{MS}}$
\end{tabular}
  & $\gamma_{1_{L,L}}^c$
  & $-\frac{4}{9} \left( 202 + 26 \frac{\pi^2}{6} - 16 n_f \right)$
\\
\hline
Ciuchini {\it et al.}~\cite{Ciuchini96b}, Buchalla~\cite{Buchalla97a}
  & $\gamma_{21}^{(0)}$ & $\gamma_{0_{S,L}}^c$ & $\frac{4}{3}$ \\
Ciuchini {\it et al.}~\cite{Ciuchini96b}, Buchalla~\cite{Buchalla97a}
  & $\gamma_{22}^{(0)}$ & $\gamma_{0_{S,S}}^c$ & $-\frac{8}{3}$ \\
Ciuchini {\it et al.}~\cite{Ciuchini96b}, Buchalla~\cite{Buchalla97a}
  & $d_1$ & $p_{0,L}^c$ & $\frac{ \hat{\gamma}_{0_{L,L}}^c }{ 2 b_0
}$ \\
Ciuchini {\it et al.}~\cite{Ciuchini96b}, Buchalla~\cite{Buchalla97a}
  & $d_2$ & $p_{0,S}^c$ & $\frac{ \hat{\gamma}_{0_{S,S}}^c }{ 2 b_0
}$ \\
Ciuchini {\it et al.}~\cite{Ciuchini96b}, Buchalla~\cite{Buchalla97a}
  & $-J$ & $p_{1,L}^c$ & $p_{0,L}^c
                          \left( \frac{ \gamma_{1_{L,L}}^c }
                                      { \gamma_{0_{L,L}}^c }
                               - \frac{ b_1 }{ b_0 } \right)$ \\
\hline
Duncan {\it et
al.}~\cite{Duncan94c,Ji91a,Brdhrst91a,Brdhrst92a,Gimenez92a}
  & $\begin{array}{c} \gamma_0 \\ \gamma_1 \end{array}$
  & $\begin{array}{c} \gamma_{0,A}^c \\ \gamma_{1,A}^c \end{array}$
  & $\begin{array}{c} -4 \\ \frac{-254}{9}
                          - \frac{56 \pi^2}{27}
                          + \frac{20 n_f}{9} \end{array}$ \\
\hline
Buras {\it et al.}~\cite{Buras90a}
  & $ \gamma^{0} $
  & $\gamma_{0_{L,L}}^f$ & 4 \\
Buras {\it et al.}~\cite{Buras90a}
  & $ \gamma^{1} $
  & $\gamma_{1_{L,L}}^f$
  & $\left( -7 + \frac{4}{9} n_f \right)$ \\
& & $p_{0,A}^f$ & 0 \\
& & $p_{1,A}^f$ & 0 \\
\end{tabular}
\end{center}
\caption{\label{t:the_ps}
Anomalous dimensions as defined by various groups and used here.  The
$p$'s are defined in Eq.~(\protect\ref{eq:the_ps}).  All the results
have been calculated using the naive dimensional regularization
scheme.}
\end{table}

We will now discuss the matching of the continuum-static theory to
the lattice-static theory.  The relevant perturbative calculations
have been done by Flynn {\it et al.}~\cite{Flynn91a}.  We want to
calculate the full theory at $m_b^\star$:
\begin{eqnarray}
\left<{\cal O}^{f}\!\left(\!m_b^\star\!\right)\right>
& = & C_L^{fc}\!\left(\!m_b^\star;q^\star\!\right)
      \left( Z_L^{cl}\!\left(\!q^\star; a \!\right)
      \left<{\cal O}_L^l\!\left(\!a\!\right)\right>
           + Z_R^{cl}\!\left(\!q^\star; a \!\right)
      \left<{\cal O}_R^l\!\left(\!a\!\right)\right>
           + Z_N^{cl}\!\left(\!q^\star; a \!\right)
      \left<{\cal O}_N^l\!\left(\!a\!\right)\right>
             \right)
    + \nonumber \\ & &
    + C_S^{fc}\!\left(\!m_b^\star;q^\star\!\right)
      Z_S^{cl}\!\left(\!q^\star; a\!\right)
      \left<{\cal O}_S^l\!\left(\!a\!\right)\right>
\end{eqnarray}
where $Z^{cl}\!\left(q^\star; a\right)$ relates the bare
lattice-static theory matrix element to the renormalized
continuum-static theory matrix element.  After linearizing the
product $C^{fc}\!\left(m_b^\star;q^\star\right)
Z^{cl}\!\left(q^\star; a\right)$ and allowing a separate coupling for
continuum-static ($\alpha_{s}^{c}$) and for lattice-static
($\alpha_{s}^{l}$) we find:
\begin{eqnarray}
\left<{\cal O}^{f}\!\left(m_b^\star\right)\right>
& = & \left[ \left( \frac{ \alpha_s^{c}\!\left(\!m_b^\star\!\right) }
                         { \alpha_s^{c}\!\left(\!q^\star\!\right) }
                    \right)^{p_{0,L}^{c}}
             \left( 1 + \frac{
\alpha_{s}^{c}\!\left(\!m_b^\star\!\right)
                             -
\alpha_{s}^{c}\!\left(\!q^\star\!\right) }
                             { 4 \pi }
                         p_{1,L}^{c}
                       + \frac{
\alpha_{s}^{c}\!\left(\!m_b^\star\!\right) }
                              { 4 \pi }
                         \left( - 14 \right)
                         \right. \right. \nonumber \\ & & \left.
\phantom{     \frac{ \alpha_s^{c}\!\left(\!m_b^\star\!\right) }
                   { \alpha_s^{c}\!\left(\!q^\star\!\right)
}^{p_{0,L}^{c}}
              ( 1 )( 1 + ) } \left.
                       + \frac{
\alpha_{s}^{l}\!\left(\!q^\star\!\right) }
                              { 4 \pi }
                         \left( 4 \ln\!\left({q^\star}^{2} a^{2}
\right)
                              - 21.7 \right)
                         \right)
      \right. \nonumber \\ & & \left.
    - \frac{1}{4}
      \left( \left( \frac{ \alpha_s^c\!\left(\!m_b^\star\!\right) }
                         { \alpha_s^c\!\left(\!q^\star\!\right) }
                    \right)^{p_{0,L}^c}
           - \left( \frac{ \alpha_s^c\!\left(\!m_b^\star\!\right) }
                         { \alpha_s^c\!\left(\!q^\star\!\right) }
                    \right)^{p_{0,S}^c}  \right)
      \frac{ \alpha_{s}^{c}\!\left(\!m_b^\star\!\right) }{4\pi} (-8)
\right]
      \left<{\cal O}_L^{l}\!\left(a\right)\right>
      \nonumber \\ & &
    + \left( \frac{ \alpha_s^{c}\!\left(\!m_b^\star\!\right) }
                  { \alpha_s^{c}\!\left(\!q^\star\!\right) }
             \right)^{p_{0,L}^{c}}
      \left\{ \frac{ \alpha_{s}^{l}\!\left(\!q^\star\!\right) }{4\pi}
             (-1.61) \left<{\cal O}_R^{l}\!\left(a\right)\right>
           + \frac{ \alpha_{s}^{l}\!\left(\!q^\star\!\right) }{4\pi}
             (-14.4) \left<{\cal O}_N^{l}\!\left(a\right)\right>
\right\}
      \nonumber \\ & &
    + \left( \frac{ \alpha_s^{c}\!\left(\!m_b^\star\!\right) }
                  { \alpha_s^{c}\!\left(\!q^\star\!\right) }
             \right)^{p_{0,S}^{c}}
      \frac{ \alpha_{s}^{c}\!\left(\!m_b^\star\!\right) }{4\pi}
      (-8) \left<{\cal O}_S^{l}\!\left(a\right)\right>
      \label{eq:fourorg}
\end{eqnarray}
where we have updated the results of Flynn {\it et
al.}~\cite{Flynn91a} by including
$(U^T)_{S,L}^c$~\cite{Ciuchini96b,Buchalla97a}, by choosing the
convention that the static-light two-point function be fit to the $A
{\rm e}^{-mt}$ model~\cite{Ukqcd96a,Eichten90c}, and by including
tadpole improvement~\cite{Hernandez94a}.

Throughout this paper we assume the convention that the $f_{B}^{}$
decay constant is extracted from the heavy-light correlators using
the model $A e^{-m t}$.  Using this model changes the heavy-quark
wave-function renormalization integral, denoted $e$, to a reduced
value $e^{(R)}$ (see Eichten and Hill~\cite{Eichten90c}).  As
mentioned by the UKQCD collaboration~\cite{Ukqcd96a}, this changes
the $D_L$=$-65.5$ of Flynn {\it et al.}\ (the additive constant in
the matching of the continuum-static ${\cal O}_{L}^{c}$ operator to
the lattice operator) to $D_L^{(R)}$=$-38.9$.  However, $e^{(R)}$
also appears in $Z_{A}^{cl}$; thus, the final values for the
coefficients of the $B$ parameters are independent of this choice if
the ratio is linearized in $\alpha^c$ and $\alpha^l$.  In addition,
any tadpole-improvement effects alter the three-point function by
twice as much as each two-point function; linearizing the ratio
cancels these effects exactly.  However, when considering the
three-point function and two-point function separately, one ought to
include the effects of tadpole improvement.  This changes the
$D_L^{(R)}$=$-38.9$ to $D_L^{(R,tad)}$=$-21.7$, as in
Eq.~(\ref{eq:fourorg}).  The large perturbative factors of the
wave-function renormalization largely cancel in the expression for
the $B_{B}^{}$ parameters.

To calculate the coefficients of $B_{B}^{}$, the renormalization
coefficient of the axial current in the static approximation is
required
\cite{Ji91a,Brdhrst91a,Brdhrst92a,Gimenez92a,Eichten90c,Eichten90b};
we linearized the results quoted in Duncan~{\it et
al.}~\cite{Duncan94c}:
\begin{eqnarray}
Z_{A}
& \equiv
    & C_{A}^{fc}\!\left(\!m_b^\star; q^\star\!\right)
      Z_{A}^{cl}\!\left(\!q^\star; a\!\right)
      \label{eq:zee_a} \\
& = & \left( \frac{ \alpha_s^{c}\!\left(\!m_b^\star\!\right) }
                  { \alpha_s^{c}\!\left(\!q^\star\!\right) }
             \right)^{p_{0,A}^{c}}
      \left( \!
             1 + \frac{ \alpha_{s}^{c}\!\left(\!m_b^\star\!\right)
                      - \alpha_{s}^{c}\!\left(\!q^\star\!\right) }{ 4
\pi }
                 p_{1,A}^{c}
               + \frac{ \alpha_{s}^{c}\!\left(\!m_b^\star\!\right) }{
4 \pi }
                 \left( \! - \frac{8}{3} \right)
               + \frac{ \alpha_{s}^{l}\!\left(\!q^\star\!\right) }{ 4
\pi }
                 \left( 2 \ln\!\left( \! {q^\star}^{2} a^{2} \!
\right)
                      - 18.59 \right)
                 \! \right)  \nonumber
\end{eqnarray}
where the $-18.59$ is from using the $e^{(R)}$ mentioned above as
well as including tadpole improvement.  If tadpole improvement had
not been used, then this value would be $-27.16$.  If $e$ had been
used instead of $e^{(R)}$, then this value would be $-40.44$.  As
long as one is consistent between Eqs.~(\ref{eq:fourorg})
and~(\ref{eq:zee_a}), these effects cancel out of the linearized
result for $B_B^{}$.

The perturbative coefficients for the $B_{B}^{}$ parameter can be
obtained by dividing the four-fermion results by the square of
$Z_{A}$ and expanding the expressions linearly in $\alpha_{s}$.
\begin{mathletters}
\label{eq:b_at_mstar}
\begin{eqnarray}
B_{B}^{}\left( m_b^\star \right)
& = & \left[ \left( \frac{ \alpha_s^{c}\!\left(m_b^\star\right) }
                         { \alpha_s^{c}\!\left(q^\star\right) }
                    \right)^{p_{0,L}^{c} - 2 p_{0,A}^{c}}
             \left( 1
                  + \frac{ \alpha_{s}^{c}\!\left(m_b^\star\right)
                         - \alpha_{s}^{c}\!\left(q^\star\right) }{ 4
\pi }
                    \left( p_{1,L}^{c} - 2 p_{1,A}^{c} \right)
                  + \frac{ \alpha_{s}^{c}(m_b^\star) }{4 \pi}
                    \left( -\frac{26}{3} \right)
                    \right. \right. \nonumber \\ & & \left.
\phantom{
      \left( \frac{ \alpha_s^{c}\!\left(m_b^\star\right) }
                  { \alpha_s^{c}\!\left(q^\star\right) }
             \right)^{ p_{0,L}^{c} - 2 p_{0,A}^{c} }
      ( 1 + ) } \left.
                  + \frac{ \alpha_{s}^{l}(q^\star) }{4 \pi}
                    \left( 15.41 \right)
                    \right)
      \right. \nonumber \\ & & \left.
    - \frac{1}{4}
      \left( \left( \frac{ \alpha_s^c\!\left(m_b^\star\right) }
                         { \alpha_s^c\!\left(q^\star\right) }
                    \right)^{p_{0,L}^c - 2 p_{0,A}^{c}}
           - \left( \frac{ \alpha_s^c\!\left(m_b^\star\right) }
                         { \alpha_s^c\!\left(q^\star\right) }
                    \right)^{p_{0,S}^c - 2 p_{0,A}^{c}}  \right)
      \frac{ \alpha_{s}^{c}\!\left(m_b^\star\right) }{4\pi} (-8)
\right]
      B_{L}
      \nonumber \\ & &
    + \left( \frac{ \alpha_s^{c}\!\left(m_b^\star\right) }
                  { \alpha_s^{c}\!\left(q^\star\right) }
             \right)^{ p_{0,L}^{c} - 2 p_{0,A}^{c} }
      \left[ \frac{ \alpha_{s}^{l}(q^\star) }{4 \pi} (- 1.61) B_{R}
           + \frac{ \alpha_{s}^{l}(q^\star) }{4 \pi} (-14.4 ) B_{N}
             \right]
      \nonumber \\ & &
    + \left( \frac{ \alpha_s^{c}\!\left(m_b^\star\right) }
                  { \alpha_s^{c}\!\left(q^\star\right) }
             \right)^{ p_{0,S}^{c} - 2 p_{0,A}^{c} }
      \frac{ \alpha_{s}^{c}(m_b^\star) }{4 \pi} (- 8 ) B_{S} \\
\nonumber \\
B_{B}^{}( m_b^\star )
& \equiv & Z_{B_L} B_{L} + Z_{B_R} B_{R}
         + Z_{B_N} B_{N} + Z_{B_S} B_{S}
\label{eq:finalB}
\end{eqnarray}
\end{mathletters}
$\!\!{}$where $Z_{B_X} = {\rm Lin}(Z_X/Z_A^2) = {\rm Lin}\left(
(C_X^{fc} Z_X^{cl}) / (C_A^{fc} Z_A^{cl})^2 \right)$, $X$ is one of
$\{L,R,N,S\}$, and ``Lin'' signifies that the ratio is linearized as
explained later in Sec.~\ref{Sec:SysErr}.  The wave-function
normalization factors of the quarks cancel between the numerator and
denominator; no ``tadpole'' factors are required for this calculation
if the coefficients are linearized.  We also note from the values in
Table~\ref{t:the_ps} that $p_{0,L}^{c} - 2 p_{0,A}^{c}$ is
identically zero.  However, $p_{0,S}^{c} - 2 p_{0,A}^{c}$ and
$p_{1,L}^{c} - 2 p_{1,A}^{c}$ are not.  If Eq.~(\ref{eq:finalB}) were
expanded into explicit $\ln(\mu/m_b)$ terms, then to first order the
perturbative matching coefficients would not contain any logs.

To calculate numerical values of the coefficients, we choose values
for the scales $\mu_b$ and $\mu$, and for the couplings
$\alpha_{s}^{c}$ and $\alpha_{s}^{l}$.  For $\alpha_s^l$, we use
$\alpha_{V}^{}$, the coupling introduced by Lepage and
Mackenzie~\cite{Lepage93}.  We use the plaquette value $-\ln W_{11} =
0.5214$ at $\beta=6.0$.  In the quenched approximation ($n_f=0$),
\begin{equation} \label{a13(20)}
-\ln (W_{11})
  =   \frac{4 \pi}{3} \alpha_{V}^{} \!\left(\frac{3.41}{a}\right)
      \left( 1
           - \alpha_{V}^{} \! \left(\frac{3.41}{a}\right)
             ( 1.19 ) \right)
\end{equation}
which uses a lattice coupling which evolves with the form
\begin{equation} \label{eq:alpha_two_loop}
\alpha_{s}(\mu)
  =   \left[ \beta_0 \ln\!\left(\frac{\mu^2}{\Lambda^2}\right)
           + \frac{\beta_1}{\beta_0}
\ln\!\left(\ln\!\left(\frac{\mu^2}{\Lambda^2}\right)\right)
             \right]^{-1}
\end{equation}
where the $\beta$ are defined in Eq.~(\ref{eq:betas}).
Eq.~(\ref{a13(20)}) defines $\alpha_{V}^{}$ and gives $\Lambda_V^{} a
= 0.169$.

Because the continuum-to-lattice matching is known only to one loop,
these perturbative expressions are sensitive in principle to the
value of the scale used in the matching.  This dependence can only be
reduced by calculating higher-order loops.  However, Lepage and
Mackenzie~\cite{Lepage93} have described a plausible procedure for
determining the scale and they have successfully tested this method
for a number of quantities.

The Lepage-Mackenzie scale $q^{\star}$ is obtained from
\begin{eqnarray}
\left< \ln\!\left(q a\right)^2 \right>
& = & \frac{ \int d^{4} q f(q) \, \ln(q^{2}) }
           { \int d^{4} q f(q) } \\
\label{eq:LMscale}
q^\star a
& = &
\exp\!\left(\frac{1}{2}\left<\ln\!\left(qa\right)^2\right>\right)
\end{eqnarray}
where $f(q)$ is the finite integrand of the lattice graphs; note that
$f(q)$ is defined by assuming that all the perturbative expressions
are expanded linearly in the coupling.  We used the integrands of
Flynn~{\it et al.}~\cite{Flynn91a}.  (These have been confirmed by
Borrelli and Pittori~\cite{Borrelli92a}.)  In Table~\ref{tb:Bscale},
we show the value of the scale for several operators.  Our value for
the scale for the static-light axial current, $q^{\star}a = 2.18$,
agrees with the calculation by Hern\'andez and
Hill~\cite{Hernandez94a}.

The Lepage-Mackenzie scales for the individual operators in
Table~\ref{tb:Bscale} are all around 2.0; however, the combined
operator for $B_{{\cal O}_{L}^{\rm full}}$ has a lower scale of 1.22.
(The scale quoted for $B_{{\cal O}_{L}^{\rm full}}$ in the original
preprint and conference proceeding~\cite{Christensen97a} was
incorrect.)  Morningstar~\cite{Morningstar94a} also found very low
scales for the perturbative renormalization of the quark mass in
NRQCD (also see the comments by Sloan~\cite{Sloan95a}); though this
could be related to renormalon effects~\cite{Sachrajda96a}.  Using
the scale of 1.22 gave large perturbative corrections.  The
Lepage-Mackenzie scale-setting procedure could be confused by taking
the ratio of matrix elements of two operators that are approximately
the same (obviously it would be inappropriate for the case of two
equal operators because $f$ would be identically zero).  We chose to
use the scale of $2.18/a$ as this is a typical scale for both $A_\mu$
and ${\cal O}_L^{full}$.

We used $\Lambda^{(5)}_{\rm QCD}$=0.175\,GeV from Duncan {\it et
al.}~\cite{Duncan94c}.  They chose values for $a^{-1}$ obtained from
the charmonium system due to the low systematic errors.  Although
they do not quote a value for $a^{-1}$ at $\beta$=6.0, they did
extrapolate $\Lambda_V^{}$ from $a^{-1}$ at $\beta$=5.7, 5.9, and 6.1
in order to find $a^{-1}$ at $\beta$=6.3.  We used this idea to
interpolate to $a^{-1}$=2.1\,GeV for $\beta$=6.0.  We also used their
method for calculating $m_b^\star$; however, our number differs
slightly because of the difference between the form of
Eq.~(\ref{eq:alpha_two_loop}) and
\begin{equation}
\alpha_s(\mu)
  =  \frac{ 1 }
          { \beta_0  \ln \!\left( \frac{ \mu^2 }{ \Lambda^2 } \right)
}
     \left[ 1
          - \frac{ \beta_1 }{ \beta_0^2 }
            \frac{ \ln \!
                   \left( \ln \!
                          \left( \frac{ \mu^2 }{ \Lambda^2 } \right)
                          \right) }
                 {  \ln \!\left( \frac{ \mu^2 }{ \Lambda^2 } \right)
}
            \right]
\end{equation}
With the full-to-continuum scale set as $\mu_b = m_b^\star =
4.33$\,GeV and the continuum-to-lattice scale set by $\mu a = q^\star
a = 2.18$, we find $\alpha^c_{s}(m_b^\star) = 0.21$ and
$\alpha^l_{s}(q^\star) = 0.18$.

Using a Monte Carlo technique, we estimated the error on the static
$B_{B}^{}$ parameter due to varying the values of the parameters used
in the perturbation theory.  A sample of one thousand was generated
using uniform deviates for the renormalization scale, lattice
spacing, the continuum $\Lambda_{\rm QCD}^{}$, and the bottom quark
mass.  The central value for each ``input''-parameter distribution
was set equal to our best value, based on those used in
references~\protect\cite{Duncan94c,Hernandez94a}.  Rather than assume
that the input parameters are known to three significant figures, we
took up to 20\% of this value to be the standard deviation for each
input parameter.  The final results were sorted numerically and the
68\% error range was taken as the ``output'' error.  This procedure
should produce more accurate estimates of errors than naive error
analysis.  Table~\ref{tb:Bcoefferr} shows the resulting error in the
coefficients.  Table~\ref{tb:resultsperturberror} shows the
corresponding error in $B_B^{}$.  The $B_{B}^{}$ parameter is very
insensitive to rather large changes in these parameters.  Variations
of 20\% in these parameters change the $B_{B}^{}$ parameter by less
than the statistical bootstrap errors.  It is particularly important
that the results are not sensitive to the lattice spacing because
there are a wide range of possible lattice spacings that could have
been used: $a^{-1} = 1.94$\,GeV from the string
tension~\cite{Bali93a}, $a^{-1} = 2.3$\,GeV from the $\rho$
mass~\cite{Bhattacharya95a}, and $a^{-1} = 2.4$\,GeV from Upsilon
spectroscopy~\cite{Davies94b}.

%
%
\begin{table}
\begin{center}
\begin{tabular}{r|ccccc}
& $A_{\mu}$ & ${\cal O}_{L}$  & ${\cal O}_{L}^{\rm full}$
& $B_{{\cal O}_{L}}$ & $B_{{\cal O}_{L}^{\rm full}}=B_{B}^{}$ \\
\hline
$q^{\star}a$ \phantom{000000}
& 2.18     & 2.01     & 2.15                 & 2.45        & 1.22
\end{tabular}
\end{center}
\caption{\label{tb:Bscale}
Renormalization scales determined by the Lepage-Mackenzie
prescription for the axial-vector current $A_{\mu}^{}$, for the raw
lattice operator ${\cal O}_{L}^{}$, and for ${\cal O}_{L}^{\rm
full}$, are similar.  Using this prescription for a ratio of matrix
elements (as for $B_B^{}$) is unstable, as described in the text;
therefore, we choose $2.18$ as the scale appropriate for $B_B^{}$.}
\end{table}
%
%
%
\begin{table}
\begin{center}
\begin{tabular}{rccccc}
 & $q^{\star}a$ & $a^{-1}$     & $m_b^\star$           &
$\Lambda_{\rm QCD}^{(5)}$ & All\\
 & $2.18$   & $2.1\,{\rm GeV}$ & $4.33\,{\rm GeV}$ & $0.175\,{\rm
GeV}$  \\
\hline
\multicolumn{6}{c}{$Z_{B_L}=1.070$} \\
\hline
10\%    & \begin{tabular}{l} $+{0.002}$ \\ $-{0.002}$ \end{tabular}
        & \begin{tabular}{l} $+{0.003}$ \\ $-{0.004}$ \end{tabular}
        & \begin{tabular}{l} $+{0.003}$ \\ $-{0.003}$ \end{tabular}
        & \begin{tabular}{l} $+{0.0008}$ \\ $-{0.0005}$ \end{tabular}
        & \begin{tabular}{l} $+{0.004}$ \\ $-{0.005}$ \end{tabular}
\\
20\%    & \begin{tabular}{l} $+{0.005}$ \\ $-{0.003}$ \end{tabular}
        & \begin{tabular}{l} $+{0.006}$ \\ $-{0.009}$ \end{tabular}
        & \begin{tabular}{l} $+{0.006}$ \\ $-{0.005}$ \end{tabular}
        & \begin{tabular}{l} $+{0.0019}$ \\ $-{0.0009}$ \end{tabular}
        & \begin{tabular}{l} $+{0.008}$ \\ $-{0.009}$ \end{tabular}
\\
\hline
\multicolumn{6}{c}{$Z_{B_R}=-0.0225$} \\
\hline
10\%    & \begin{tabular}{l} $+{0.0005}$ \\ $-{0.0006}$ \end{tabular}
        & -  & - & -
        & \begin{tabular}{l} $+{0.0005}$ \\ $-{0.0006}$ \end{tabular}
\\
20\%    & \begin{tabular}{l} $+{0.0009}$ \\ $-{0.0015}$ \end{tabular}
        & - & - & -
        & \begin{tabular}{l} $+{0.0009}$ \\ $-{0.0015}$ \end{tabular}
\\
\hline
\multicolumn{6}{c}{$Z_{B_N}=-0.202$} \\
\hline
10\%    & \begin{tabular}{l} $+{0.005}$ \\ $-{0.006}$ \end{tabular}
        & - & - & -
        & \begin{tabular}{l} $+{0.005}$ \\ $-{0.006}$ \end{tabular}
\\
20\%    & \begin{tabular}{l} $+{0.008}$ \\ $-{0.012}$ \end{tabular}
        & - & - & -
        & \begin{tabular}{l} $+{0.008}$ \\ $-{0.012}$ \end{tabular}
\\
\hline
\multicolumn{6}{c}{$Z_{B_S}=-0.137$} \\
\hline
10\%    & - & -
        & \begin{tabular}{l} $+{0.003}$ \\ $-{0.003}$ \end{tabular}
        & \begin{tabular}{l} $+{0.002}$ \\ $-{0.003}$ \end{tabular}
        & \begin{tabular}{l} $+{0.003}$ \\ $-{0.004}$ \end{tabular}
\\
20\%    & - & -
        & \begin{tabular}{l} $+{0.006}$ \\ $-{0.005}$ \end{tabular}
        & \begin{tabular}{l} $+{0.005}$ \\ $-{0.007}$ \end{tabular}
        & \begin{tabular}{l} $+{0.006}$ \\ $-{0.008}$ \end{tabular}
\\
\end{tabular}
\end{center}
\caption{\label{tb:Bcoefferr}
The absolute changes from our preferred values of the coefficients
$Z_{B_L}$, $Z_{B_R}$, $Z_{B_N}$, and $Z_{B_S}$ as the parameters
$q^{\star}a$, $a^{-1}$, $m_b^\star$, and $\Lambda_{\rm QCD}^{(5)}$,
are varied by 10\% and 20\% first individually, and then jointly
(``All''), from our preferred values.  We do not imply and need not
assume that the input parameters are known to three significant
figures (indeed the coefficients are quite insensitive to 20\%
variations in the values of the parameters); rather, we chose central
values based on references~\protect\cite{Duncan94c,Hernandez94a}.}
\end{table}
%
%
%
\begin{table}
\begin{center}
\begin{tabular}{rccccc}
& $\kappa=0.152$ & $\kappa=0.154$ & $\kappa=0.155$
                 & $\kappa=0.156$ & $\kappa_c=0.157$ \\
\hline
10\%    & \begin{tabular}{l} $+{0.007}$ \\ $-{0.009}$ \end{tabular}
        & \begin{tabular}{l} $+{0.007}$ \\ $-{0.009}$ \end{tabular}
        & \begin{tabular}{l} $+{0.007}$ \\ $-{0.009}$ \end{tabular}
        & \begin{tabular}{l} $+{0.007}$ \\ $-{0.009}$ \end{tabular}
        & \begin{tabular}{l} $+{0.007}$ \\ $-{0.009}$ \end{tabular}
\\
20\%    & \begin{tabular}{l} $+{0.013}$ \\ $-{0.017}$ \end{tabular}
        & \begin{tabular}{l} $+{0.013}$ \\ $-{0.017}$ \end{tabular}
        & \begin{tabular}{l} $+{0.013}$ \\ $-{0.017}$ \end{tabular}
        & \begin{tabular}{l} $+{0.013}$ \\ $-{0.017}$ \end{tabular}
        & \begin{tabular}{l} $+{0.013}$ \\ $-{0.017}$ \end{tabular}
\end{tabular}
\end{center}
\caption{\label{tb:resultsperturberror}
The absolute changes in $B_{B}^{}$, from
Eq.~(\protect\ref{eq:fit-combine}), due to changes in the
coefficients $Z_{B_L}$, $Z_{B_R}$, $Z_{B_N}$, and $Z_{B_S}$ as the
parameters $q^{\star}a$, $a^{-1}$, $m_b^\star$, and $\Lambda_{\rm
QCD}^{(5)}$ are varied jointly by 10\% and 20\% from our preferred
values. }
\end{table}

To compare the results of $B_{B}^{}$ parameters, in the next section
we list our results in terms of the one-loop and the two-loop
renormalization-group-invariant parameter~\cite{Buras90a}.  We also
scale $B_{B}^{}$ down to 2.0\,GeV for the comparison to some other
groups (Sec.~\ref{Sec:Compare}) which is discussed later.

To compare at one-loop, we scaled $B_B^{}$ and calculated
$\widehat{B}_B$ using
\begin{eqnarray}
 B_B \!\left( \mu_1 \right)
& = & \left( \frac{  \alpha_s \!\left( \mu_1 \right) }
                  {  \alpha_s \!\left( \mu_2 \right) }
             \right)^{ p_{0,L}^f - 2 p_{0,A}^f }
       B_B \!\left( \mu_2 \right)
       \label{eq:Bhat_running} \\
\widehat{B}_B
& = &  \alpha_s \!\left( \mu_2 \right)^{ -\left( p_{0,L}^f - 2
p_{0,A}^f
                                                 \right)}
       B_B \!\left( \mu_2 \right)
       \label{eq:B_running}
\end{eqnarray}
where $p$ is defined in Eq.~(\ref{eq:the_ps}) with the relevant
anomalous dimensions listed in Table~\ref{t:the_ps}.  Since this is a
one-loop calculation, we used
\begin{eqnarray}
\alpha_s^{-1}\!\left(\mu\right)
& = & \beta_0  \ln \!\left( \left( \frac{ \mu }{ \Lambda } \right)^2
\right)
\label{eq:alpha_one_loop}
\end{eqnarray}

Although a one-loop calculation is traditional, one can also
calculate a two-loop renormalization-group-invariant
$\widehat{B}_{B}^{}$ parameter since the required perturbative
calculations have been done.
\begin{eqnarray}
 B_B \!\left( \mu_1 \right)
& = & \left( \frac{  \alpha_s \!\left( \mu_1 \right) }
                  {  \alpha_s \!\left( \mu_2 \right) }
             \right)^{ p_{0,L}^f - 2 p_{0,A}^f }
      \left( 1
           + \frac{  \alpha_s \!\left( \mu_1 \right)
                  -  \alpha_s \!\left( \mu_2 \right) }{ 4\pi }
             \left( p_{1,L}^f - 2 p_{1,A}^f \right) \right)
       B_B \!\left( \mu_2 \right) \\
\widehat{B}_B
& = &  \alpha_s \!\left( \mu_2 \right)^{ -\left( p_{0,L}^f - 2
p_{0,A}^f
                                                 \right) }
      \left( 1
           - \frac{  \alpha_s \!\left( \mu_2 \right) }{ 4\pi }
             \left( p_{1,L}^f - 2 p_{1,A}^f \right) \right)
       B_B \!\left( \mu_2 \right)
\end{eqnarray}
Again $p$ is defined in Eq.~(\ref{eq:the_ps}) and the relevant
anomalous dimensions are listed in Table~\ref{t:the_ps}.
Eq.~(\ref{eq:alpha_two_loop}) was used to scale $B_B^{}$ and
calculate $\widehat{B}_B$ to second order.

In making a comparison to other groups, one can use either $B_B^{}$
evaluated at some scale or $\widehat{B}_B^{}$.  There are
disadvantages to both.  For the former, either a common scale needs
to be agreed upon or $B_B^{}$ must be scaled.  For the latter, the
dependence of $\widehat{B}_B^{}$ on the choice of $n_f$ and $\Lambda$
is not negligible; $\widehat{B}_B^{}$ can vary by as much as 4\% to
5\% (see Sec.~\ref{Sec:SysErr}).  This dependence is also relevant to
using one-loop versus two-loop because the difference in the value of
$\Lambda^{(n_f)}_{}$ between Eqs.~(\ref{eq:alpha_one_loop})
and~(\ref{eq:alpha_two_loop}) can vary by as much as 10\%.  The
advantage to comparing $B_B^{}$ at some scale is that the dependence
on $n_f$ and $\Lambda$ is less significant ($\approx 1\%$, see
Table~\ref{tb:resultsperturberror}).  Also, given a value for
$B_B^{}\!\left(m_b^\star\right)$, one can quote a value for
$\widehat{B}_B^{}$ using either 4 or 5 flavors since $m_b^\star$ is
the boundary between $n_f$ = 4 and 5 flavors.  These give different
constant values of $\widehat{B}_B^{}$ for the different flavor
regimes.  One should be explicit about which is quoted.

Even though the numerical results are for the quenched theory, we use
$n_{\!f}$=$5$ for \mbox{$\mu \geq m_b^\star$} and $n_{\!f}$=$4$ for
\mbox{$\mu \leq m_b^\star$}.  There is some evidence from studies of
the QCD coupling that the effects of omitting dynamical fermions can
be modeled by using the correct number of flavors in the $\beta$
function (see Sloan~\cite{Sloan95a} for a review).

\section{Systematic errors in the matching}
\label{Sec:SysErr}

The discussion until now has not revealed any large systematic errors
in the perturbative matching that could explain the difference
between our result and UKQCD's.  In this section we investigate the
systematic error caused by combining the perturbative coefficients
for the two-point and three-point functions in different ways to form
the matching coefficient for the $B_B^{}$ operator.  The UKQCD
collaboration found a $20\%$ effect when they changed the way they
organized their perturbative coefficients~\cite{Ukqcd96a}.

We consider three different ways of calculating the coefficients
$Z_{B_X}$ to consider these effects.  For convenience, we define the
following, where $X$ is one of $\{ L, R, N, S, A \}$:
\begin{eqnarray}
Z_X
& \equiv & {\rm product\ of\ } \left( C_X^{fc} Z_X^{cl} \right)
      \nonumber \\
& = & \left( \frac{ \alpha_s^{c}\!\left(\!m_b^\star\!\right) }
                  { \alpha_s^{c}\!\left(\!q^\star\!\right) }
             \right)^{p_{0,X}^{c}} \!\!
      \left( 1 + \frac{ \alpha_{s}^{c}\!\left(\!m_b^\star\!\right)
                      - \alpha_{s}^{c}\!\left(\!q^\star\!\right) }{ 4
\pi }
                 p_{1,X}^{c}
               + \frac{ \alpha_{s}^{c}\!\left(\!m_b^\star\!\right) }{
4 \pi }
                 \left( D_X^c \right)
                 \right)
\phantom{just for spacing}
      \nonumber \\  & &
\phantom{
      \left( \frac{ \alpha_s^{c}\!\left(\!m_b^\star\!\right) }
                  { \alpha_s^{c}\!\left(\!q^\star\!\right) }
             \right)
    } \times
      \left( 1 + \frac{ \alpha_{s}^{l}\!\left(\!q^\star\!\right) }{ 4
\pi }
                 \left( d_X^l \ln\!\left(q^\star a \right)^{2} +
D_X^l \right)
                 \right)
      \label{eq:prod} \\
{\rm Lin}(Z_X)
& \equiv & {\rm linearization\ of\ } \left( C_X^{fc} Z_X^{cl} \right)
      \nonumber \\
& = & \left( \frac{ \alpha_s^{c}\!\left(\!m_b^\star\!\right) }
                  { \alpha_s^{c}\!\left(\!q^\star\!\right) }
             \right)^{p_{0,X}^{c}} \!\!
      \left( 1 + \frac{ \alpha_{s}^{c}\!\left(\!m_b^\star\!\right)
                      - \alpha_{s}^{c}\!\left(\!q^\star\!\right) }{ 4
\pi }
                 p_{1,X}^{c}
               + \frac{ \alpha_{s}^{c}\!\left(\!m_b^\star\!\right) }{
4 \pi }
                 \left( D_X^c \right)
                 \nonumber \right. \\ & &
\phantom{
      \left( \frac{ \alpha_s^{c}\!\left(\!m_b^\star\!\right) }
                  { \alpha_s^{c}\!\left(\!q^\star\!\right) }
             \right)^{p_{0,X}^{c}} \!\!
    } \left. \phantom{(1)}
               + \frac{ \alpha_{s}^{l}\!\left(\!q^\star\!\right) }{ 4
\pi }
                 \left( d_X^l \ln\!\left(q^\star a \right)^{2} +
D_X^l \right)
                 \right)
      \label{eq:lin}
\end{eqnarray}
We wish to compare three forms of linearization: ``fully linearized''
${{\rm Lin}(Z_L/Z_A^2)}$, ``not linearized'' ${Z_L/Z_A^2}$ and
``partially linearized'' ${{\rm Lin}(Z_L)/{\rm Lin}(Z_A)^2}$.  The
UKQCD collaboration compared their ${\rm Lin}(Z_L)/{\rm Lin}(Z_A)^2$
to $Z_L/{\rm Lin}(Z_A)^2$ when they found their $20\%$ effect in
$B_B^{}$.  Since $C_A^{fc}$ is very close to 1, ${\rm Lin}(Z_A)$ is
approximately equal to $Z_A$. Thus comparing their preferred ${\rm
Lin}(Z_L)/{\rm Lin}(Z_A)^2$ to their alternative $Z_L/{\rm
Lin}(Z_A)^2$ is essentially the same as comparing ${\rm
Lin}(Z_L)/{\rm Lin}(Z_A)^2$ (partially linearized) to ${Z_L/Z_A^2}$
(not linearized).

To allow a direct comparison with others, our not-linearized results
have changed somewhat from those reported in the conference
proceedings~\cite{Christensen97a} and the original preprint of this
article which calculated the not-linearized result for $Z_R$ and
$Z_N$ as $( \alpha_s^{c}\!\left(\!m_b^\star\!\right) /
\alpha_s^{c}\!\left(\!q^\star\!\right) )^{p_{0,L}^{c}} Z_X^{cl}$
rather than $C_L^{fc} Z_X^{cl}$.

In Table~\ref{t:linear} we show the coefficients of the individual
$B_B^{}$ parameters for the three different linearizations described,
both with and without tadpole improvement.  Table~\ref{tb:Bours}
shows the corresponding value for $\widehat{B}_B$ at both one-loop
and at two-loops. The variation among the three different
linearizations of the non-tadpole-improved coefficients is much
larger than for the tadpole-improved coefficients.  Because there are
equal numbers of quarks in the numerator and denominator, the
individual $B_B^{}$ parameters should be independent of the
wave-function normalization of both the heavy and the light quarks.
This implies that the coefficients should be independent of tadpole
improvement.  Tables~\ref{t:linear} and~\ref{tb:Bours} show that this
is only true for the fully-linearized quantity, ${{\rm
Lin}(Z_L/Z_A^2)}$.
\begin{table}[p]
\begin{center}
\begin{tabular}{l|rrrr|r}
Method & $Z_{B_L}$ & $Z_{B_R}$ & $Z_{B_N}$ & $Z_{B_S}$ &
$B_B(m_b^\star)$ \\
\hline \multicolumn{6}{c}{With Tadpole Improvement} \\ \hline
Lin$(Z_X/Z_A^2)$        & $ 1.070^{+0.009}_{-0.009}$
                        & $-0.022^{+0.001}_{-0.001}$
                        & $-0.202^{+0.008}_{-0.012}$
                        & $-0.137^{+0.006}_{-0.008}$
                        & $ 0.98^{+0.04}_{-0.04}$ \\
$(Z_X/Z_A^2)$           & $ 1.066^{+0.020}_{-0.016}$
                        & $-0.031^{+0.002}_{-0.004}$
                        & $-0.275^{+0.021}_{-0.037}$
                        & $-0.246^{+0.013}_{-0.025}$
                        & $ 0.96^{+0.04}_{-0.04}$ \\
Lin$(Z_X)$/Lin$(Z_A)^2$ & $ 1.003^{+0.014}_{-0.021}$
                        & $-0.041^{+0.003}_{-0.005}$
                        & $-0.371^{+0.026}_{-0.050}$
                        & $-0.251^{+0.015}_{-0.021}$
                        & $ 0.80^{+0.04}_{-0.04}$ \\
\hline \multicolumn{6}{c}{Without Tadpole Improvement} \\ \hline
Lin$(Z_X/Z_A^2)$        & $ 1.070^{+0.009}_{-0.008}$
                        & $-0.022^{+0.001}_{-0.001}$
                        & $-0.202^{+0.008}_{-0.012}$
                        & $-0.137^{+0.006}_{-0.007}$
                        & $ 0.98^{+0.04}_{-0.04}$ \\
$(Z_X/Z_A^2)$           & $ 1.030^{+0.011}_{-0.014}$
                        & $-0.043^{+0.004}_{-0.007}$
                        & $-0.384^{+0.036}_{-0.065}$
                        & $-0.343^{+0.018}_{-0.042}$
                        & $ 0.87^{+0.04}_{-0.04}$ \\
Lin$(Z_X)$/Lin$(Z_A)^2$ & $ 0.802^{+0.039}_{-0.082}$
                        & $-0.059^{+0.005}_{-0.010}$
                        & $-0.529^{+0.049}_{-0.092}$
                        & $-0.358^{+0.025}_{-0.044}$
                        & $ 0.49^{+0.04}_{-0.04}$
\end{tabular}
\end{center}
\caption{\label{t:linear}
The effects of different linearizations on the coefficients: The
errors on the coefficients are the statistical errors of varying the
parameters $q^{\star}a$, $a^{-1}$, $m_b^\star$, and $\Lambda_{\rm
QCD}^{(5)}$ by 20\% from our preferred values.  The error bars on
$B_B\!\left(m_b^\star\right)$ are the bootstrap errors.
$B_B\!\left(m_b^\star\right)$ is the chiral extrapolation of the
``combine-then-fit'' values from Eq.~(\protect\ref{eq:combine-fit}).}
\end{table}
\begin{table}
\begin{center}
\begin{tabular}{lcc|cc|cc}
       &           &       & \multicolumn{2}{c|}{one-loop}
                           & \multicolumn{2}{c}{two-loop}  \\
Method & $B(4.33)$ & $n_f$ & $\Lambda$ & $\widehat{B}_B$
                           & $\Lambda$ & $\widehat{B}_B$ \\
\hline \multicolumn{7}{c}{With Tadpole Improvement} \\ \hline
Lin$(Z_X/Z_A^2)$ & 0.98(4)
 & \begin{tabular}{c}    5    \\    4    \end{tabular}
 & \begin{tabular}{c}   175   \\   226   \end{tabular}
 & \begin{tabular}{c} 1.40(6) \\ 1.36(6) \end{tabular}
 & \begin{tabular}{c}   175   \\   246   \end{tabular}
 & \begin{tabular}{c} 1.50(6) \\ 1.46(6) \end{tabular} \\
$(Z_X/Z_A^2)$  & 0.96(4)
 & \begin{tabular}{c}    5    \\    4    \end{tabular}
 & \begin{tabular}{c}   175   \\   226   \end{tabular}
 & \begin{tabular}{c} 1.37(6) \\ 1.33(6) \end{tabular}
 & \begin{tabular}{c}   175   \\   246   \end{tabular}
 & \begin{tabular}{c} 1.47(6) \\ 1.43(6) \end{tabular} \\
Lin$(Z_X)$/Lin$(Z_A)^2$ & 0.80(4)
 & \begin{tabular}{c}    5    \\    4    \end{tabular}
 & \begin{tabular}{c}   175   \\   226   \end{tabular}
 & \begin{tabular}{c} 1.14(6) \\ 1.11(6) \end{tabular}
 & \begin{tabular}{c}   175   \\   246   \end{tabular}
 & \begin{tabular}{c} 1.23(6) \\ 1.19(6) \end{tabular} \\
\hline \multicolumn{7}{c}{Without Tadpole Improvement} \\ \hline
Lin$(Z_X/Z_A^2)$ & 0.98(4)
 & \begin{tabular}{c}    5    \\    4    \end{tabular}
 & \begin{tabular}{c}   175   \\   226   \end{tabular}
 & \begin{tabular}{c} 1.40(6) \\ 1.36(6) \end{tabular}
 & \begin{tabular}{c}   175   \\   246   \end{tabular}
 & \begin{tabular}{c} 1.50(6) \\ 1.46(6) \end{tabular} \\
$(Z_X/Z_A^2)$  & 0.87(4)
 & \begin{tabular}{c}    5    \\    4    \end{tabular}
 & \begin{tabular}{c}   175   \\   226   \end{tabular}
 & \begin{tabular}{c} 1.24(6) \\ 1.21(6) \end{tabular}
 & \begin{tabular}{c}   175   \\   246   \end{tabular}
 & \begin{tabular}{c} 1.34(6) \\ 1.30(6) \end{tabular} \\
Lin$(Z_X)$/Lin$(Z_A)^2$ & 0.49(4)
 & \begin{tabular}{c}    5    \\    4    \end{tabular}
 & \begin{tabular}{c}   175   \\   226   \end{tabular}
 & \begin{tabular}{c} 0.70(6) \\ 0.68(6) \end{tabular}
 & \begin{tabular}{c}   175   \\   246   \end{tabular}
 & \begin{tabular}{c} 0.75(6) \\ 0.73(6) \end{tabular}
\end{tabular}
\end{center}
\caption{\label{tb:Bours}
{}From the $B_B^{}\!\left(m_b^\star\right)$ result extracted by Monte
Carlo, listed in Table~\protect\ref{t:linear}, we calculated a
$\widehat{B}_B$ with both 4 and 5 flavors (see text).  The
Lin$(Z_X/Z_A^2)$ results are our preferred values.  As mentioned in
the text, $\widehat{B}_B$ varies with $n_f$ and $\Lambda^{(n_f)}$ as
well as with loop-order.}
\end{table}

{}From Table~\ref{t:linear}, the overall change in $B_B(m_b^\star)$
for
the three different linearizations, when calculated with the
tadpole-improved coefficients, is $20\%$.  However, when calculated
from non-tadpole-improved perturbative coefficients, $B_B(m_b^\star)$
can change by a much larger factor.  This suggests that the
order-$\alpha^2$ effects may be large.  While these can be treated in
a variety of ways, we think that they can be treated well or treated
poorly.  For example, the use of tadpole improvement stabilizes the
central values and reduces statistical errors.  The UKQCD
collaboration did not use tadpole improvement, which suggests that
their perturbative coefficients may be unnecessarily sensitive to
their choice of linearization. (Their preferred choice is what we
call ``partially linearized''; they also considered what we call
``not linearized''.)  Their decision not to use tadpole-improvement
was forced upon them by the way they implemented the light-quark
field rotations which were required to remove $O(a)$ corrections to
matrix elements~\cite{Ukqcd93a}.

We rank the various organizations of perturbation theory in
decreasing order of preference: fully linearized, not linearized,
partially linearized.  We discuss, in turn, several (related)
disadvantages with partial linearizing: larger relative statistical
errors, increased sensitivity to the value of the lattice coupling
constant (via choice of prescription), and non-optimal handling of
order-$\alpha^2$ terms.  Firstly, due to the larger off-diagonal
coefficients in the terms of the sum in Eq.~(\ref{eq:finalB}), the
numerical result for $B_B^{}\!\left(m_b^\star\right)$ using
non-tadpole-improved partially-linearized coefficients (the last row
of Table~\ref{t:linear}) has a larger relative statistical error than
do the results from the other choices of linearization.

Secondly, we studied the stability of the results from three groups:
the $\beta = 6.2$ clover-static UKQCD simulation~\cite{Ukqcd96a}, the
$\beta = 6.0$ clover-static Gim\'enez \& Martinelli
simulation~\cite{Gimenez97a}, and our $\beta = 6.0$ Wilson-static
simulation (both tadpole-improved and not-tadpole improved).  All
three of these groups that have done static $B_B$ simulations used
slightly different ways of evaluating the perturbative coefficients.
We have analyzed all simulation data consistently to facilitate
comparisons of the results.  We compared the linearizations for two
lattice couplings ($\tilde{\alpha}$ and
$\alpha_{V}^{}(q^*a\!=\!2.18)$) and for summing the logarithms (\`a
la renormalization-group (RG) techniques) {\sl versus\/} not summing
the logarithms.  These are discussed further in
Sec.~\ref{Sec:Compare}.  We see the same trends in each group's data.
Each group's partially-linearized result is less stable under
variations of $\alpha$ than is their not linearized or fully
linearized.  Their fully-linearized result is close to their
not-linearized result; these are 20\% higher than their
partial-linearized result.

Thirdly, we believe that partial linearization does a poor job of
organizing higher-order terms.  The treatment of $O(\alpha^2)$ terms
in partially-linearized coefficients causes the low values seen by
all groups by linearizing some terms but not the whole ratio.  We
prefer the fully-linearized method because it removes all of these
$O(\alpha^2)$ terms (as in a Taylor expansion) by linearizing the
whole ratio.  Fully- or not-linearizing the coefficients treats the
$O(\alpha^2)$ terms more appropriately than does partially
linearizing.

Our preferred choice of linearization (full) can also be motivated by
the non-perturbative renormalization method, introduced by the
Rome-Southampton group~\cite{Martinelli94a}. The non-perturbative
renormalization method for $B_B$ parameter would be very similar to
that used to obtain the renormalization constants for the kaon $B$
parameter~\cite{Donini95a,Donini96a}, in which all the factors of the
lattice wave function normalization of the quarks cancel explicitly
for the $B$ parameter. In perturbation theory, this corresponds to
our preferred full linearization. The non-perturbative method only
determines the lattice part of the renormalization factor; a choice
of linearization would still have to be made for the continuum
factor. However, the continuum factor can and should be calculated to
next to leading order~\cite{Martinelli94a}, making it less sensitive
to the different choices of linearizations.

In summary, our preference for the treatment of the coefficients is
to fully linearize the ratio (in the notation of this section,
$Z_{B_L}$ is ${{\rm Lin}(Z_L/Z_A^2)}$).  This gives a result which
has no order-$\alpha^2$ terms, which is insensitive to the inclusion
of tadpole improvement and to the wave-function normalization model
by allowing explicit cancelations, and which reduces the statistical
errors in $B_B^{}$.  The quantitative consequences of our choice are
discussed in the following section where we compare the results of
different groups.

Just as the numerical value of $B_{B}^{}$ is stable because of
cancelations of correlated fluctuations in numerator and denominator,
we have argued that so too are its perturbative corrections when
fully linearized.  The fully-linearized perturbatively-calculated
coefficients for $B_{B}^{}$ are likely more reliable than those for
the product $B_{B}^{} f_{B}^{2}$, the quantity which is required in
the analysis of $\overline{B^{0}}${\bf --}$B^{0}$ mixing experiments.
In the Appendix, we discuss our recommendation for how to linearize
the product $B_B f_B^{2}$.

\section{World Comparison}
\label{Sec:Compare}

In Table~\ref{tb:Bothers}~(\ref{tb:Bothers_two}), we show a
collection of results from several groups scaled to give
$B_{B}^{}(m_b^\star)$, $B_{B}^{}(2.0\,{\rm GeV})$, and the one-loop
(two-loop) renormalization-group-invariant $\widehat{B}_{B}$
parameter.  Results from both static and relativistic-quark
simulations are shown.  The simulations using relativistic heavy
Wilson quarks~\cite{Soni96a,Bernard88a,Abada92a,Jlqcd96a} calculate
the $B_{B}^{}$ parameter for quark masses around charm and
extrapolate up to the physical mass, using a fit model of the form
\begin{equation}
B_{B} = B^{0}_{B} + \frac{ B^{1}_{B}} {M}
\label{eq:Bmodelmass}
\end{equation}
The value of $B^{0}_{B}$ should be the same as the static theory
result.  (It is better to do a combined analysis of relativistic and
static quarks to obtain a value for $B_{B}^{}$.)  We call $B^{0}_{B}$
the ``extrapolated-static'' value.
\begin{table}
\begin{center}
\begin{tabular}{lcccl|cclll}
 & & & $\mu_2$ & & & $\Lambda$ & \multicolumn{3}{c}{one-loop} \\
Method & Ref. & $\beta$ &  (GeV)  & $B(\mu_2)$
              & $n_f$ & (MeV) & $B(2.0)$ & $B(4.33)$ &
$\widehat{B}_B$ \\
\hline
Static-Clover & \cite{Ukqcd96a}
 & 6.2 & $m_b$=5.0 & {\sl 0.69(4)}
 & \begin{tabular}{c}        5      \\    4    \end{tabular}
 & \begin{tabular}{c}       130     \\   200   \end{tabular}
 & \begin{tabular}{l}    \ \ \ -    \\ 0.75(4) \end{tabular}
 & \begin{tabular}{l}    \ \ \ -    \\ 0.70(4) \end{tabular}
 & \begin{tabular}{l} {\sl 1.02(6)} \\ 0.98(6) \end{tabular} \\
\cline{6-10}
Static-Clover & \cite{Ukqcd96a}
 & 6.2 & $m_b$=5.0 &      0.81(4)
 & \begin{tabular}{c}        5      \\    4    \end{tabular}
 & \begin{tabular}{c}       130     \\   200   \end{tabular}
 & \begin{tabular}{l}    \ \ \ -    \\ 0.87(4) \end{tabular}
 & \begin{tabular}{l}    \ \ \ -    \\ 0.82(4) \end{tabular}
 & \begin{tabular}{l} {\sl 1.19(6)} \\ 1.14(6) \end{tabular} \\
\cline{6-10}
Static-Clover & \cite{Gimenez97a}
 & 6.0 & $m_b$=5.0 & {\sl 0.54(4)}
 & \begin{tabular}{c}        5      \\    4    \end{tabular}
 & \begin{tabular}{c}       151     \\   200   \end{tabular}
 & \begin{tabular}{l}    \ \ \ -    \\ 0.59(4) \end{tabular}
 & \begin{tabular}{l}    \ \ \ -    \\ 0.55(4) \end{tabular}
 & \begin{tabular}{l} {\sl 0.79(6)} \\ 0.77(6) \end{tabular} \\
\cline{6-10}
Static-Clover & \cite{Gimenez97a}
 & 6.0 & $m_b$=5.0 & {\sl 0.76(5)}
 & \begin{tabular}{c}        5      \\    4    \end{tabular}
 & \begin{tabular}{c}       151     \\   200   \end{tabular}
 & \begin{tabular}{l}    \ \ \ -    \\ 0.82(5) \end{tabular}
 & \begin{tabular}{l}    \ \ \ -    \\ 0.77(5) \end{tabular}
 & \begin{tabular}{l} {\sl 1.11(7)} \\ 1.08(7) \end{tabular} \\
\hline
Static-Wilson & \begin{tabular}{c} this \\ work \end{tabular}
 & 6.0 & $m_b^\star$=4.33 & {\sl 0.98(4)}
 & \begin{tabular}{c}   5      \\   4     \end{tabular}
 & \begin{tabular}{c}  175     \\  226    \end{tabular}
 & \begin{tabular}{l}\ \ \ -   \\ 1.05(4) \end{tabular}
 &               {\sl 0.98(4)}
 & \begin{tabular}{l} 1.40(6)  \\ 1.36(6) \end{tabular} \\
\hline
Extrap. Static & \cite{Soni96a}
 & 5.7-6.3 & $\mu$=2.0 & {\sl 1.04(5)}
 & \begin{tabular}{c}    4    \\    4    \end{tabular}
 & \begin{tabular}{c}   200   \\   226   \end{tabular}
 &               {\sl 1.04(5)}
 & \begin{tabular}{l} 0.97(5) \\ 0.97(5) \end{tabular}
 & \begin{tabular}{l} 1.36(7) \\ 1.34(6) \end{tabular} \\
\cline{6-10}
Extrap. Static & \cite{Abada92a}
 & 6.4 & $\mu$=3.7 & {\sl 0.90(5)}
 & \begin{tabular}{c}    0    \\    4    \end{tabular}
 & \begin{tabular}{c}   200   \\   200   \end{tabular}
 & \begin{tabular}{l} 0.94(5) \\ 0.95(5) \end{tabular}
 & \begin{tabular}{l} 0.89(5) \\ 0.89(5) \end{tabular}
 & \begin{tabular}{l} 1.21(7) \\ 1.25(7) \end{tabular} \\
\hline
Wilson-Wilson & \cite{Soni96a}
 & 5.7-6.3 & $\mu$=2.0 & {\sl 0.96(6)}
 & \begin{tabular}{c} 4 \\ 4 \end{tabular}
 & \begin{tabular}{c} 200 \\ 226 \end{tabular}
 &               {\sl 0.96(6)}
 & \begin{tabular}{l} 0.90(6) \\ 0.89(6) \end{tabular}
 & \begin{tabular}{l} 1.25(8) \\ 1.24(8) \end{tabular} \\
\cline{6-10}
Wilson-Wilson & \cite{Soni96a,Bernard88a}
 & 6.1 & $\mu$=2.0 & {\sl 1.01(15)}
 & \begin{tabular}{c} 4 \\ 4 \end{tabular}
 & \begin{tabular}{c} 200 \\ 226 \end{tabular}
 &               {\sl 1.01(15)}
 & \begin{tabular}{l} 0.94(13) \\ 0.94(14) \end{tabular}
 & \begin{tabular}{l} 1.32(20) \\ 1.30(19) \end{tabular} \\
\cline{6-10}
Wilson-Wilson &  \cite{Jlqcd96a}
 & 6.1 & $m_b$=5.0 & {\sl 0.895(47)}
 & \begin{tabular}{c}    0    \\    4    \\    5    \end{tabular}
 & \begin{tabular}{c}   239   \\   239   \\   183   \end{tabular}
 & \begin{tabular}{l} 0.96(5) \\ 0.98(5) \\\ \ \ -  \end{tabular}
 & \begin{tabular}{l} 0.90(5) \\ 0.91(5) \\\ \ \ -  \end{tabular}
 & \begin{tabular}{l} 1.21(6) \\ 1.25(7) \\ 1.29(7) \end{tabular} \\
\cline{6-10}
Wilson-Wilson & \cite{Jlqcd96a}
 & 6.3 & $m_b$=5.0 & {\sl 0.840(60)}
 & \begin{tabular}{c}    0    \\    4    \\    5    \end{tabular}
 & \begin{tabular}{c}   246   \\   246   \\   189   \end{tabular}
 & \begin{tabular}{l} 0.90(6) \\ 0.92(6) \\\ \ \ -  \end{tabular}
 & \begin{tabular}{l} 0.85(6) \\ 0.85(6) \\\ \ \ -  \end{tabular}
 & \begin{tabular}{l} 1.14(8) \\ 1.17(8) \\ 1.20(9) \end{tabular} \\
\cline{6-10}
Wilson-Wilson & \cite{Abada92a}
 & 6.4 & $\mu$=3.7 & {\sl 0.86(5)}
 & \begin{tabular}{c}         0    \\    4    \end{tabular}
 & \begin{tabular}{c}        200   \\   200   \end{tabular}
 & \begin{tabular}{l}      0.90(5) \\ 0.91(5) \end{tabular}
 & \begin{tabular}{l}      0.85(5) \\ 0.85(5) \end{tabular}
 & \begin{tabular}{l} {\sl 1.16(7)}\\ 1.19(7) \end{tabular} \\
\hline
Sum Rule & \cite{Narison94a}
 &  & $m_b$=4.6 & {\sl 1.00(15)}
 & \begin{tabular}{c}     5    \\    4     \end{tabular}
 & \begin{tabular}{c}    175   \\   227    \end{tabular}
 & \begin{tabular}{l} \ \ \ -  \\ 1.08(16) \end{tabular}
 & \begin{tabular}{l} \ \ \ -  \\ 1.00(15) \end{tabular}
 & \begin{tabular}{l} 1.43(22) \\ 1.39(21) \end{tabular} \\
\end{tabular}
\end{center}
\caption{\label{tb:Bothers}
The authors' numbers, quoted at the listed value for $\mu_2$, have
been scaled using Eq.~(\protect\ref{eq:Bhat_running}) to
$\mu$=2.0\,GeV and to $m_b^\star$=4.33\,GeV.  The {\sl slanted\/}
numbers are those that the cited authors quote.  We calculated
$B_B^{}\!\left(m_b^\star\right)$ in the Static-Wilson case and then
scaled it to 2.0\,GeV using $n_{\!f}$=4 and calculated a
$\widehat{B}_B$ with both 4 and 5 flavors (see text).  The value
quoted by this work uses the fully-linearized tadpole-improved
coefficients.  The JLQCD collaboration cite their $\Lambda$'s as
$n_{\!f}$=0 values.  When Abada {\it et al.}\ quotes a
$\widehat{B}_B$ for the Wilson quarks, they use $n_{\!f}$=0.  We
scaled both groups' results using both $n_{\!f}$=0 and $n_{\!f}$=4.}
\end{table}
\begin{table}
\begin{center}
\begin{tabular}{lcccl|cclll}
 & & & $\mu_2$ & & & $\Lambda$ & \multicolumn{3}{c}{two-loop} \\
Method & Ref. & $\beta$ &  (GeV)  & $B(\mu_2)$
              & $n_f$ & (MeV) & $B(2.0)$ & $B(4.33)$ &
$\widehat{B}_B$ \\
\hline
Static-Clover & \cite{Ukqcd96a}
 & 6.2 & $m_b$=5.0 & {\sl 0.69(4)}
 & \begin{tabular}{c}     5   \\    4    \end{tabular}
 & \begin{tabular}{c}    130  \\   200   \end{tabular}
 & \begin{tabular}{l} \ \ \ - \\ 0.74(5) \end{tabular}
 & \begin{tabular}{l} \ \ \ - \\ 0.70(4) \end{tabular}
 & \begin{tabular}{l} 1.09(6) \\ 1.05(6) \end{tabular} \\
\cline{6-10}
Static-Clover & \cite{Ukqcd96a}
 & 6.2 & $m_b$=5.0 &      0.81(4)
 & \begin{tabular}{c}        5      \\    4    \end{tabular}
 & \begin{tabular}{c}       130     \\   200   \end{tabular}
 & \begin{tabular}{l}    \ \ \ -    \\ 0.86(4) \end{tabular}
 & \begin{tabular}{l}    \ \ \ -    \\ 0.81(4) \end{tabular}
 & \begin{tabular}{l}      1.27(6)  \\ 1.23(6) \end{tabular} \\
\cline{6-10}
Static-Clover & \cite{Gimenez97a}
 & 6.0 & $m_b$=5.0 & {\sl 0.54(4)}
 & \begin{tabular}{c}        5      \\    4    \end{tabular}
 & \begin{tabular}{c}       136     \\   200   \end{tabular}
 & \begin{tabular}{l}    \ \ \ -    \\ 0.58(4) \end{tabular}
 & \begin{tabular}{l}    \ \ \ -    \\ 0.54(4) \end{tabular}
 & \begin{tabular}{l} {\sl 0.86(6)} \\ 0.82(6) \end{tabular} \\
\cline{6-10}
Static-Clover & \cite{Gimenez97a}
 & 6.0 & $m_b$=5.0 & {\sl 0.76(5)}
 & \begin{tabular}{c}        5      \\    4    \end{tabular}
 & \begin{tabular}{c}       136     \\   200   \end{tabular}
 & \begin{tabular}{l}    \ \ \ -    \\ 0.77(5) \end{tabular}
 & \begin{tabular}{l}    \ \ \ -    \\ 0.81(5) \end{tabular}
 & \begin{tabular}{l} {\sl 1.21(8)} \\ 1.16(8) \end{tabular} \\
\hline
Static-Wilson & \begin{tabular}{c} this \\ work \end{tabular}
 & 6.0 & $m_b^\star$=4.33 & {\sl 0.98(4)}
 & \begin{tabular}{c}   5      \\   4     \end{tabular}
 & \begin{tabular}{c}  175     \\  246    \end{tabular}
 & \begin{tabular}{l}\ \ \ -   \\ 1.04(4) \end{tabular}
 &               {\sl 0.98(4)}
 & \begin{tabular}{l} 1.50(6)  \\ 1.46(6) \end{tabular} \\
\hline
Extrap. Static & \cite{Soni96a}
 & 5.7-6.3 & $\mu$=2.0 & {\sl 1.04(5)}
 & \begin{tabular}{c}    4    \\    4    \end{tabular}
 & \begin{tabular}{c}   200   \\   246   \end{tabular}
 &               {\sl 1.04(5)}
 & \begin{tabular}{l} 0.98(5) \\ 0.98(5) \end{tabular}
 & \begin{tabular}{l} 1.49(7) \\ 1.46(7) \end{tabular} \\
\cline{6-10}
Extrap. Static & \cite{Abada92a}
 & 6.4 & $\mu$=3.7 & {\sl 0.90(5)}
 & \begin{tabular}{c}    0    \\    4    \end{tabular}
 & \begin{tabular}{c}   200   \\   200   \end{tabular}
 & \begin{tabular}{l} 0.93(5) \\ 0.94(5) \end{tabular}
 & \begin{tabular}{l} 0.89(5) \\ 0.89(5) \end{tabular}
 & \begin{tabular}{l} 1.29(7) \\ 1.35(7) \end{tabular} \\
\hline
Wilson-Wilson & \cite{Soni96a}
 & 5.7-6.3 & $\mu$=2.0 & {\sl 0.96(6)}
 & \begin{tabular}{c} 4 \\ 4 \end{tabular}
 & \begin{tabular}{c} 200 \\ 246 \end{tabular}
 &               {\sl 0.96(6)}
 & \begin{tabular}{l} 0.91(6) \\ 0.90(6) \end{tabular}
 & \begin{tabular}{l} 1.37(9) \\ 1.35(9) \end{tabular} \\
\cline{6-10}
Wilson-Wilson & \cite{Soni96a,Bernard88a}
 & 6.1 & $\mu$=2.0 & {\sl 1.01(15)}
 & \begin{tabular}{c} 4 \\ 4 \end{tabular}
 & \begin{tabular}{c} 200 \\ 246 \end{tabular}
 &               {\sl 1.01(15)}
 & \begin{tabular}{l} 0.96(14) \\ 0.95(14) \end{tabular}
 & \begin{tabular}{l} 1.44(21) \\ 1.42(21) \end{tabular} \\
\cline{6-10}
Wilson-Wilson &  \cite{Jlqcd96a}
 & 6.1 & $m_b$=5.0 & {\sl 0.895(47)}
 & \begin{tabular}{c}    0    \\    4    \\    5    \end{tabular}
 & \begin{tabular}{c}   239   \\   239   \\   183   \end{tabular}
 & \begin{tabular}{l} 0.94(5) \\ 0.96(5) \\\ \ \ -  \end{tabular}
 & \begin{tabular}{l} 0.90(5) \\ 0.90(5) \\\ \ \ -  \end{tabular}
 & \begin{tabular}{l} 1.29(7) \\ 1.35(7) \\ 1.36(7) \end{tabular} \\
\cline{6-10}
Wilson-Wilson & \cite{Jlqcd96a}
 & 6.3 & $m_b$=5.0 & {\sl 0.840(60)}
 & \begin{tabular}{c}    0    \\    4    \\    5    \end{tabular}
 & \begin{tabular}{c}   246   \\   246   \\   189   \end{tabular}
 & \begin{tabular}{l} 0.88(6) \\ 0.90(6) \\\ \ \ -  \end{tabular}
 & \begin{tabular}{l} 0.85(6) \\ 0.85(6) \\\ \ \ -  \end{tabular}
 & \begin{tabular}{l} 1.21(9) \\ 1.26(9) \\ 1.30(9) \end{tabular} \\
\cline{6-10}
Wilson-Wilson & \cite{Abada92a}
 & 6.4 & $\mu$=3.7 & {\sl 0.86(5)}
 & \begin{tabular}{c}         0    \\    4    \end{tabular}
 & \begin{tabular}{c}        200   \\   200   \end{tabular}
 & \begin{tabular}{l}      0.89(5) \\ 0.90(5) \end{tabular}
 & \begin{tabular}{l}      0.85(5) \\ 0.85(5) \end{tabular}
 & \begin{tabular}{l} {\sl 1.24(7)}\\ 1.29(7) \end{tabular} \\
\hline
Sum Rule & \cite{Narison94a}
 &  & $m_b$=4.6 & {\sl 1.00(15)}
 & \begin{tabular}{c}     5    \\    4     \end{tabular}
 & \begin{tabular}{c}    175   \\   227    \end{tabular}
 & \begin{tabular}{l} \ \ \ -  \\ 1.07(16) \end{tabular}
 & \begin{tabular}{l} \ \ \ -  \\ 1.00(15) \end{tabular}
 & \begin{tabular}{l} 1.54(23) \\ 1.50(22) \end{tabular} \\
\end{tabular}
\end{center}
\caption{\label{tb:Bothers_two}
This table repeats the analysis in Table~\protect\ref{tb:Bothers},
using the {\it two-loop\/} renormalization group invariant $B_B^{}$
parameter.}
\end{table}

Tables~\ref{tb:Bothers} and~\ref{tb:Bothers_two} show that values for
$B_{B}^{}$ obtained from Wilson action simulations are basically
consistent; the small differences can be explained by small
lattice-spacing and finite-volume effects.  Our result is consistent
with that of Bernard and Soni, as reported by Soni~\cite{Soni96a},
for the extrapolated static Wilson fermions.

Since the original appearance of the preprint for this article ({\tt
hep-lat/9610026}, version 1), data has been made available which
allows a more detailed comparison between ourselves (on the high end
of the world results) and others (on the low end).  Firstly, we have
added the updated numbers from Gim\'enez \&
Martinelli~\cite{Gimenez97a} to Tables~\ref{tb:Bothers}
and~\ref{tb:Bothers_two}.  Secondly, we note that
Wittig~\cite{Wittig97a} has a nice review on the subject of leptonic
decays of lattice heavy quarks, in which he compares the results of
UKQCD~\cite{Ukqcd96a}, Gim\'enez \& Martinelli~\cite{Gimenez97a} and
the preprint of this article.

In his Sec.~4.2, Wittig offers Table 9 for comparison, using our
non-tadpole-improved results.  We find that the tadpole-improved
Wilson results improve the non-tadpole-improved results, so we prefer
to compare their clover-improved results to our tadpole-improved
results.  Our analogous comparison results in the numbers listed in
Table~\ref{tb:Wittig}.
\begin{table}
\begin{center}
\begin{tabular}{c|cccc}
$B_B(4.33\,{\rm GeV})$
& UKQCD~\protect{\cite{Ukqcd96a}}
& G\&M~\protect{\cite{Gimenez97a}}
& {\sc tad}
& {\sc no-tad}  \\ \hline
fl {\scriptsize (M3)} \begin{tabular}{cc}
$\Sigma$ & \begin{tabular}{c} $\tilde{\alpha}$ \\
                              $\alpha_{V}^{}$ \end{tabular} \\
\hline $\Sigma\!\!\!\!/$
         & \begin{tabular}{c} $\tilde{\alpha}$ \\
                              $\alpha_{V}^{}$ \end{tabular}
			    \end{tabular}
         & \begin{tabular}{c} 0.84(5) \\ 0.83(5) \\
                              0.75(5) \\ 0.77(5) \end{tabular}
         & \begin{tabular}{c} 0.85(4) \\ 0.85(3) \\
                              0.84(4) \\ 0.84(3) \end{tabular}
         & \begin{tabular}{c} 0.97(4) \\ 0.97(4) \\
                              0.96(4) \\ 0.96(4) \end{tabular}
         & \begin{tabular}{c} 0.97(4) \\ 0.97(4) \\
                              0.96(4) \\ 0.96(4) \end{tabular} \\
\hline
nl {\scriptsize (M1)} \begin{tabular}{cc}
$\Sigma$ & \begin{tabular}{c} $\tilde{\alpha}$ \\
                              $\alpha_{V}^{}$ \end{tabular} \\
\hline $\Sigma\!\!\!\!/$
         & \begin{tabular}{c} $\tilde{\alpha}$ \\
                              $\alpha_{V}^{}$ \end{tabular}
			    \end{tabular}
         & \begin{tabular}{c} 0.85(5) \\ 0.82(5) \\
                              0.78(5) \\ 0.76(5) \end{tabular}
         & \begin{tabular}{c} 0.84(3) \\ 0.82(3) \\
                              0.83(3) \\ 0.81(3) \end{tabular}
         & \begin{tabular}{c} 0.95(4) \\ 0.96(4) \\
                              0.94(4) \\ 0.95(4) \end{tabular}
         & \begin{tabular}{c} 0.81(4) \\ 0.87(4) \\
                              0.80(4) \\ 0.86(4) \end{tabular} \\
\hline
pl {\scriptsize (M2)} \begin{tabular}{cc}
$\Sigma$ & \begin{tabular}{c} $\tilde{\alpha}$ \\
                              $\alpha_{V}^{}$ \end{tabular} \\
\hline $\Sigma\!\!\!\!/$
         & \begin{tabular}{c} $\tilde{\alpha}$ \\
                              $\alpha_{V}^{}$ \end{tabular}
			    \end{tabular}
         & \begin{tabular}{c} 0.72(5) \\ 0.62(4) \\
                              0.60(4) \\ 0.54(4) \end{tabular}
         & \begin{tabular}{c} 0.70(3) \\ 0.62(3) \\
                              0.68(3) \\ 0.61(3) \end{tabular}
         & \begin{tabular}{c} 0.75(4) \\ 0.80(4) \\
                              0.73(4) \\ 0.78(4) \end{tabular}
         & \begin{tabular}{c} 0.30(3) \\ 0.49(4) \\
                              0.27(3) \\ 0.47(4) \end{tabular} \\
\end{tabular}
\end{center}
\caption{\label{tb:Wittig}
Comparison between the fit-then-combine
(Eq.~(\protect{\ref{eq:fit-combine}})) analysis for $B(m_b)$ of the
three groups' data.  These numbers are for $m_b\!=\!4.33\,{\rm GeV}$,
$q^\ast\!=\!2.18a^{-1}$, and $n_f\!=\!5$.  ``fl'' is
fully-linearized, ``nl'' is not-linearized, and ``pl'' is
partially-linearized.  ``M1,'' ``M2,'' and ``M3'' refer to the
notation of Gim\'enez \& Martinelli~\protect{\cite{Gimenez97a}} and
Wittig~\protect{\cite{Wittig97a}}.  We list both our tadpole-improved
({\sc tad}) and our non-tadpole-improved ({\sc no-tad}) results.  The
errors are roughly estimated from statistical errors on the raw $B_X$
values and approximate errors on the coefficients.  See the text for
comments on $\tilde{\alpha}$ and $\alpha_{V}^{}$.}
\end{table}

In the comparison, we use $n_f=5$ and $\Lambda^{(5)}_{\rm
QCD}=0.175\,{\rm GeV}$ which result in $\alpha_s^{\rm
cont}(m_b\!=\!4.33\,{\rm GeV})=0.21$.  We also use our two-loop
$\Lambda_V^{} a = 0.169$ to scale $\alpha_V^{\rm
latt}(q^\ast\!=\!2.18a^{-1})=0.18$. Both $\alpha_s$ are with
two-loops from Eq.~(\ref{eq:alpha_two_loop}).  We also compare using
$\tilde{\alpha} = 6 / (4\pi \beta u_0^4)$, which is 0.132 for the
UKQCD collaboration~\cite{Ukqcd96a}, 0.1458 for Gim\'enez \&
Martinelli~\cite{Gimenez97a}, and 0.198 for us.  (For each group, we
used $u_0=1/8\kappa_{\rm crit}$ to calculate $\tilde{\alpha}$.)  In
addition, since the original UKQCD results\footnote{UKQCD did
investigate the use of renormalization group improved perturbation
theory, but they did not use it to calculate their final results.} do
not sum the logarithms (\`a la RG techniques), Table~\ref{tb:Wittig}
lists both summing logs ($\Sigma$) and not summing the logs
($\Sigma\!\!\!/\,$).

Rather than calculate a $q^* a$ (Eq.~(\ref{eq:LMscale})) and a
$\Lambda_V^{} a$ (Eq.~(\ref{a13(20)})) for the clover action, we used
our values.  Since $\alpha_{V}^{}$ is a function of
$(q^*a/\Lambda_V^{} a)$, $\alpha_{V}^{}(q^*)$ is the same for all
three groups.  We note that $a^{-1}=2.9$ was used for UKQCD and
$a^{-1}=2.1$ was used for both Gim\'enez \& Martinelli and ourselves.
Since $q^\ast a$ was chosen to be $2.18$ for all three groups, the
scales in the comparison of Table~\ref{tb:Wittig} are different.
This is the reason that the UKQCD $\Sigma\!\!\!/\,$ results differ
from their $\Sigma$ results.  The $\Sigma\!\!\!/\,$ results are more
sensitive to the scale of the perturbative matching.

Though not listed in the table, we are able to reproduce the results
of both UKQCD~\cite{Ukqcd96a} and Gim\'enez \&
Martinelli~\cite{Gimenez97a} for $m_b\!=\!5.0\,{\rm GeV}$,
$\mu\!=\!a^{-1}$, $n_f\!=\!4$, and $\Lambda^{(4)}_{\rm
QCD}=0.200\,{\rm GeV}$ when we tailor the respective calculations
according to the method presented in each paper\footnote{To reproduce
UKQCD's~\cite{Ukqcd96a} $0.69(4)$ and $0.81$ (the latter is our
conversion of their quoted $\hat{B} = 1.19$), do not sum the logs,
use $\mu a$=$1$, and do not include the cross term, $(U^T)^c_{LS}$,
in the coefficient of ${\cal O}_L$.  To reproduce Gim\'enez' \&
Martinelli's~\cite{Gimenez97a} Table 3, sum the logs and include the
cross term, but use $\mu a$=$1$, even for the
$\alpha_{V}^{}(q^*a\!=\!2.18)$ case.}.  Also, we agree with the
results of Wittig~\cite{Wittig97a} for our $\Sigma$--$\alpha_{V}^{}$
entries when we use his parameters.

Both UKQCD's and Gim\'enez's \& Martinelli's quoted values for the
static $B_{B}^{}$ are lower than all of the other results.  One
possible reason for these low results is that they used the clover
action for the light quarks, which does not have corrections to the
continuum limit that are linear in the lattice spacing, whereas the
standard Wilson fermion action does have such artifact terms.
However, the Wilson results are stable over four different lattice
spacings, which implies that the lattice artifact terms alone cannot
account for the difference between the clover results and the Wilson
numbers.

Table~\ref{tb:Wittig} shows that the not-linearized (and
fully-linearized) static clover results for $B_{B}^{}$ are larger
than the partially-linearized results, as is discussed in the
original papers.  The clover-static results that use the
not-linearized matching are in better agreement, though still low,
with the results from simulations which use relativistic heavy quarks
to simulate the $b$ quark (see Table~\ref{tb:Bothers}).  All the
published data~\cite{Soni96a,Abada92a} on calculating $B_{B}^{}$
using relativistic heavy quarks favor a negative value of $B_{B}^{1}$
in Eq.~(\ref{eq:Bmodelmass}).  For consistency, the static value of
$B_{B}^{}$ should be higher than the value of $B_{B}^{}$ extrapolated
to the $b$ quark mass. This is true for our result and favors the
higher clover-static results.

The various choices made in the calculation have non-negligible
effects.  One can choose which action to use (Wilson {\sl vs\/}
clover), whether or not to tadpole-improve, and which linearization
method to use.  The choice between our tadpole-improved Wilson-static
action and the non-tadpole-improved clover-static action has a 15\%
effect in both the fully- and not-linearized
($\Sigma$--$\alpha_{V}^{}$) cases.  This is a 20\% effect for the
partially-linearized case.  In addition, tadpole-improvement
stabilizes the Wilson-static results to the extent that one can make
a better comparison of different linearizations between
tadpole-Wilson-static and clover-static than between
non-tadpole-Wilson-static and clover-static.  Finally, there is a
20\% effect due to choice of linearization for either action.  This
linearization effect is at least as large as the effect due to choice
of action.  For reasons given in Sec.~\ref{Sec:SysErr}, our favorite
choice of linearization is the fully-linearized treatment.

A similar trend can be seen in each group's results:
partially-linearized values are smaller and less stable than either
not-linearized or fully-linearized values.  This is due to
$O(\alpha^2)$ terms which may or may not cancel to varying degrees.
The partially-linearized treatment only linearizes part of the ratio
which causes its value to be misleadingly low.  The not-linearized
and fully-linearized treatments are better because they do not do
this.  The fully-linearized treatment is preferred because it treats
$O(\alpha^2)$ terms uniformly by removing them (as one does in an
expansion).

\section{Conclusion}
\label{Sec:Con}

Our primary result from this tadpole-improved $\beta=6.0$
Wilson-static calculation is $B_{B}^{}(m_b^\star) =
0.98^{+4}_{-4}{}^{+3}_{-18}$, where the errors are statistical
(bootstrap) and systematic, respectively.  The overall systematic
error is obtained in quadrature from the following: ${}^{+3}_{-3}$
from the choice of fit-range, ${}^{+1}_{-2}$ from the
parameter-dependence of the perturbative-calculated mixing
coefficients, and ${}^{+0}_{-18}$ due to the the choice of
linearization of the coefficients, as was discussed in
Sec.~\ref{Sec:SysErr}.  The unusual asymmetry of the latter
systematic error reflects our preference for a particular choice of
linearization (``full'').  Our second favorite choice
(``not-linearized'') results in a central value of $0.96$.  We quote
a very conservative systematic error to encompass our least favorite
choice ($0.80$ from ``partial linearization'') even though we have
argued against this choice.  Systematic errors from finite lattice
spacing and from quenching are not estimated.

Tables~\ref{tb:Bothers} and~\ref{tb:Bothers_two} show that values for
$B_{B}^{}$ obtained from Wilson action simulations are basically
consistent; the small differences can be explained by small
lattice-spacing and finite-volume effects.  The simulations all favor
a negative value of $B_{B}^{1}$ in
Eq.~(\ref{eq:Bmodelmass})~\cite{Soni96a}.  For consistency, this
implies that the static value of $B_{B}^{}$ should be higher than the
value of $B_{B}^{}$ extrapolated to the $b$ quark mass.  Our number
is on the high end of the comparison in Table~\ref{tb:Bothers} and is
consistent with that of Bernard and Soni~\cite{Soni96a} who use
extrapolated static Wilson fermions.

In Sec.~\ref{Sec:SysErr} we investigated the effect of changing the
way the four-fermion operator renormalization and the axial-current
renormalization were combined to form the matching coefficient for
the $B_B^{}$ parameter.  We presented arguments that suggested that
our preferred way of organizing the continuum-to-lattice matching
(full linearization) was superior to any other method we considered.
We also showed that making a different choice could lower the result
by as much as 20\%.  Besides the linearizations,
Table~\ref{tb:Wittig} shows a 15\% difference due to choice of action
between our tadpole-improved $\beta=6.0$ Wilson-static and the
non-tadpole-improved $\beta=6.0$ and $6.2$ clover-static results in
the fully- and not-linearized cases.  (The Wilson results are at the
high end of the world data and the clover results are at the low
end.)  Partial-linearization leaves a 20\% effect due to choice of
action.  The effect due to choice of linearization is at least as
large as the effect due to choice of action.

Although all organizations of perturbation theory at one-loop are
theoretically equal, some are more equal than others!  Fully
linearizing gives a result which has no order-$\alpha^2$ terms and
which is insensitive to the inclusion of tadpole improvement and to
the wave-function normalization model by allowing explicit
cancelations (which reduces the statistical errors in $B_B^{}$).

In our perturbative-matching procedure we included next-to-leading
order log terms and organized the perturbative matching in a way that
we believe reduces higher-order corrections.  Also we used the
automatic scale-setting procedure of Lepage and Mackenzie to find the
``best'' scale to use in the lattice-to-continuum matching.  The
agreement of our results with relativistic heavy quark results
supports our procedure.  Our conclusion is that for the Wilson-static
case, the use of tadpole improvement and of a fully-linearized
treatment of the mixing coefficients is preferred.  Of course, this
may become less important numerically with increased coupling and/or
improved actions; however, we still recommend the procedure.

Although sensible things can be done to reduce the effects of
higher-order perturbative corrections in the lattice-to-continuum
matching, this will remain the dominant uncertainty in the
calculation of $B_B^{}$ in the static theory.  In principle, ``all''
that is required is a calculation of the two-loop anomalous dimension
of the ${\cal O}_{L}$ and $A_\mu$ operators on the lattice.  Although
this calculation is very difficult, new developments in lattice
perturbation theory for Wilson quarks~\cite{Burgio96b} and a new
stochastic way of doing lattice perturbation theory~\cite{Renzo94a}
may make these calculations more tractable in the future.  A more
immediate solution would be to use the numerical renormalization
technique, developed by the Rome-Southampton
group~\cite{Martinelli94a}, which has already been used to determine
the lattice perturbative coefficients for static
$f_{B}^{}$~\cite{Donini96a}, for the kaon $B_{K}$
parameter~\cite{Donini95a}, and for other important quantities.

The relative consistency of the Wilson $B_{B}^{}$ results motivates a
large study using both relativistic and static quarks in the same
simulations to constrain the interpolation to the $B$ mass.  To
constrain the systematic errors, the results of simulations with
different lattice spacings and volumes should be combined to take the
continuum limit.  This kind of study will also help to control the
perturbative-matching errors, as the effects of the higher-order
perturbative terms are reduced as the continuum limit is taken.  (A
nice example of this for the effects of different renormalization
prescriptions on light-quark decay constants has been given by the
GF11 group~\cite{Butler94a}.)

Once mixing in the $\overline{B_{s}^{0}}${\bf --}$B_{s}^{0}$ system
has been measured experimentally, the results can be combined with
data from $\overline{B^{0}}${\bf --}$B^{0}$ mixing experiments to
calculate the $V_{ts}/V_{td}$ ratio of CKM matrix elements.  The
advantage of calculating this ratio is that various uncertain
standard-model factors cancel.  However, a value of ${ B_{B_s}
f_{B_s}^{2} }/{ B_{B} f_{B}^{2} }$ is required.  As there are a large
number of lattice results on the calculation of
$f_{B_{s}}/f_{B}$~\cite{Allton96a}, here we concentrate on the ratio
$B_{B_s}/B_{B}$.

Using a fit model which is linear in the quark mass, we obtain
$B_{B_s}/B_{B} = 0.99^{+1}_{-1}(1)$.  (The first error is statistical
(bootstrap) and the second is the standard deviation of the fitted
value for ``reasonable choices'' of fit range.)  Even though the
ratio $B_{B_s}/B_{B}$ is determined quite precisely, it is not
resolved whether $B_{B_s}$ is greater than or less than $B_{B}$ since
the $B_{B}$ parameter is found to depend weakly on the quark mass.
Other groups~\cite{Ukqcd96a,Jlqcd96a,Bernard97a} have reported
similar findings.  Most lattice simulations have found that $f_{B_s}$
is between ten and twenty percent larger than
$f_{B}$~\cite{Allton96a}.  Bernard {\it et al.}~\cite{Bernard97a}
have extracted the ratio of $B_{s} f_{B_s}^{2} / B_{B} f_{B}^{2}$
directly by doing individual fits to the three-point function in
relativistic quark simulations.  This is a promising approach for
relativistic heavy quarks.  We did not try it because of concerns
about the signal-to-noise ratio and about the size of the
perturbative coefficients in the static theory.

Our result also contains an unknown systematic error due to
quenching.  Quenched chiral perturbation theory predicts the effects
of quenching to be small for $B_{B}^{}$~\cite{Booth95a,Sharpe95a};
this conclusion was confirmed by Bernard and Soni~\cite{Soni96a} who
calculated $B_{B}^{}$ in both quenched and dynamical simulations.  In
Soni's review~\cite{Soni96a} of the lattice calculation of weak
matrix elements at the {\sl Lattice~'95\/} conference, he quotes a
value of $B_{B}^{}(2\,{\rm GeV}) = 1.0 \pm 0.15$ (90\% confidence) as
his best estimate of the $B_{B}^{}$ parameter.  Our result,
$B_{B}^{}(2\,{\rm GeV}) = 1.05^{+4}_{-4}{}^{+3}_{-19}$, is consistent
with this value and with the vacuum-saturation-approximation value,
$1$.

\acknowledgements

This work is supported in part by the U.S.\ Department of Energy
under grant numbers DE-FG05-84ER40154 and DE-FC02-91ER75661, and by
the University of Kentucky Center for Computational Sciences.  The
computations were carried out at NERSC.

\appendix

\section*{Linearization Strategy for \mbox{$B_B f_B^{2}$}}

In the analysis of $\overline{B^{0}}${\bf --}$B^{0}$ mixing
experiments the value of $B_B f_B^{2}$ is required.  Here we discuss
the linearization options for combining $B_B$ and $f_B$ from a
variety of linearizations of these quantities.  If a not-linearized
$B_{B}^{}$ is multiplied with a not-linearized $(f_B)^2$, then the
only order-$\alpha^2$ effects which remain are due specifically to
not linearizing the numerator of $B_B^{}$.  We estimate this effect
to be on the order of 10\%.  If one multiplies a partially-linearized
$B_{B}^{}$ with a linearized $f_B$, Lin$(Z_A)^2$, then there should
be no order-$\alpha^2$ effects due to the product.  However, if one
mixes a linearized with a not-linearized $B_{B}^{}$ and $f_B$, then
there can be terms of almost 20\%.  Although the difference between
$Z_A$ and Lin$(Z_A)$ is smaller than 5\%, the difference between
Lin$(Z_A)^2$ and Lin$(Z_A^2)$ is just over 15\%.  The practical
drawback of using a $B_{B}^{}$ which is not linearized or is
partially linearized is that there are order-$\alpha^2$ terms present
which may or may not cancel when the $B_{B}^{}$ is combined with an
$f_B$.

The practical drawback to using the fully-linearized $B_{B}^{}$ is
linearizing the product $B_B^{} f_B^2$.  This is easily remedied.
The fully-linearized $B_B^{}$, $B_{fl}$, essentially has the form
\begin{equation}
B_{fl} = \left( 1 + \alpha^c A + \alpha^l C \right) B^{raw}
\end{equation}
where the $B_{R}$, $B_{N}$, and $B_{S}$ can be included by adjusting
the values of $A$ and $C$ appropriately.  When this is combined with
the square of the linearized $f$,
\begin{equation}
f_{lin} = \left( 1 + \alpha^c D + \alpha^l E \right) f^{raw}
\end{equation}
it would be convenient to get a linearized result with no
order-$\alpha^2$ terms:
\begin{equation}
\left( 1 + \alpha^c A + \alpha^l C + 2 \alpha^c D + 2 \alpha^l E
\right)
 B^{raw} \left( f^{raw}\right)^2
\label{eq:linearized}
\end{equation}
Since
\begin{equation}
\left( 1 + \alpha A \right)
\left( 1 + \alpha \frac{B}{1 + \alpha A} \right) =
\left( 1 + \alpha A + \alpha B \right)
\end{equation}
this is straightforward to accomplish.  The product of $B_{fl}$ with
the linearized square of
\begin{equation}
f_{lin}^\prime = \left( 1 + \alpha^c \frac{ D }
                                          { \left( \frac{ B_{fl} }
                                                        { B_L^{raw} }
                                                   \right) }
                          + \alpha^l \frac{ E }
                                          { \left( \frac{ B_{fl} }
                                                        { B_L^{raw} }
                                                   \right) }
                        \right)
                 f^{raw}
\end{equation}
gives Eq.~(\ref{eq:linearized}) with no order-$\alpha^2$ terms due to
coefficient multiplication.  Our $B_L^{raw}$ value can be read from
the first row of Table~\ref{tb:Bparam}.

While the product of the partially-linearized $B_{B}^{}$ with the
linearized $f_B$ also does not have any order-$\alpha^2$ terms due to
coefficient multiplication, the partially-linearized $B_{B}^{}$ by
itself has order-$\alpha^2$ terms which are on the order of 18\% (See
Tables~\ref{t:linear} and~\ref{tb:Bours}).  The advantage of our
method is that all three quantities $B_B(m_b^\star)$,
$f_B(m_b^\star)$ and $B_Bf_B^2(m_b^\star)$ have no order-$\alpha^2$
terms due to coefficient multiplication, and that $B_B(m_b^\star)$ is
stable against the inclusion of tadpole improvement and the choice of
wave-function normalization.

\pagebreak[4]



\end{document}